\documentstyle[12pt,epsf]{article}

\catcode`\@=11
\def\numberbysection{\@addtoreset{equation}{section}
        \def\theequation{\thesection.\arabic{equation}}}
\def\beq{\begin{equation}}
\def\eeq{\end{equation}}
\def\barr{\begin{eqnarray}}
\def\earr{\end{eqnarray}}

\def\disp{\displaystyle}

\def\I{{\rm Im}\ }
\def\R{{\rm Re}\ }
\def\Z{{\bf Z}}
\def\C{{\bf C}}
\newcommand{\nl}{\nonumber \\}
\newcommand{\al}{\alpha}
\newcommand{\be}{\beta}
\newcommand{\ga}{\gamma}

\newcommand{\ep}{\epsilon}

\numberbysection
\begin{document}
\begin{titlepage}
\begin{center}
\hfill  \quad DFF 288/10/97 \\
\hfill  \quad hep-th/9710248 \\
\vskip .6 in
{\LARGE Isomonodromic Properties} \\
\ \\
{\LARGE of the Seiberg-Witten Solution}
\vskip 0.2in
Andrea CAPPELLI, \ \ Paolo VALTANCOLI \ \ and \ \ Luca VERGNANO \\
\vskip 0.1in
{\em I.N.F.N. and Dipartimento di Fisica}\\
{\em Largo E. Fermi 2, I-50125 Firenze, Italy}
 \end{center}
\vskip .5 in
\begin{abstract}
The Seiberg-Witten solution of $N=2$ supersymmetric 
$SU(2)$ gauge theories with matter is analysed as an isomonodromy problem.
We show that the holomorphic section describing
the effective action can be deformed by moving its singularities on 
the moduli space while keeping their monodromies invariant.
Well-known examples of isomonodromic sections are given by the 
correlators of two-dimensional rational conformal field theories -- the
conformal blocks.
The Seiberg-Witten section similarly admits the operations
of braiding and fusing of its singularities,
which obey the Yang-Baxter and Pentagonal identities, respectively.
Using them, we easily find the complete expressions of the monodromies
with affine term, and the full quantum numbers of the BPS spectrum.
While the braiding describes the quark-monopole transmutation, 
the fusing implies the superconformal points in the moduli space.
In the simplest case of three singularities, 
the supersymmetric sections are directly related to
the conformal blocks of the logarithmic minimal models.
\end{abstract}

\vskip 1.cm
\vfill
\hfill October 1997
\end{titlepage}
\pagenumbering{arabic}
%

\section{Introduction}

The beautiful exact solution of the low-energy effective actions of
$N=2$ supersymmetric gauge theories in four dimensions \cite{s-w}
has led to dramatic developments in field theory and string theory 
over the last three years \cite{peskin}. 
In this paper, we would like to analyse some mathematical
aspects of the exact solution and compare them with analogous 
properties of the correlators of two-dimensional conformal field theory -
a well-understood exactly solvable problem \cite{cft}. 
Our general motivation is
to understand the integrable structure behind the Seiberg-Witten
solution and to develop a framework for extending the solution to other
field-theory observables. 
We shall make a few steps in this program, which we find interesting,
and, meanwhile, we shall obtain some physical results for the spectrum
of these theories. We shall consider the simple case of $N=2$
supersymmetric $SU(2)$ gauge theories with $N_f$ massive quark 
hypermultiplets, $N_f=0,1,2,3$ \cite{s-w}.

A crucial element of the $N=2$ supersymmetric solution is the
holomorphicity of the prepotential ${\cal F} ( \Psi )$ as a functional
of $N=2$ chiral superfield $\Psi$ \cite{seiberg}. 
This is a consequence of chiral 
decomposition of supersymmetric representations\footnote{
In the $N=1$ action, the superpotential ${\cal W}(\phi)$ and the gauge 
effective coupling $\tau (\phi)$ are similarly holomorphic 
in the $N=1$ chiral superfield $\phi$.}: in
formulae\footnote{
In Eq.(\ref{holo}), $\alpha=1,2$ and $i=1,2$
are the spinor and supersymmetry indices, respectively.},
\beq
\overline{D}_\alpha^i {\cal F} (\Psi) = 0  \quad{\rm \ and \ }\quad
\overline{D}_\alpha^i \Psi  = 0 \ \ \longrightarrow \ \ 
{\delta \over \delta{\overline\Psi}}  {\cal F} (\Psi) = 0 \ .
\label{holo}\eeq
Once the fields have acquired a v.e.v., $\langle \Psi \rangle =a$, 
holomorphicity in field space implies that ${\cal F} (a)$ is a 
holomorphic function of the v.e.v.,
$\partial /\partial{\overline{a}} \ {\cal F} (a) = 0 $, 
and of the coordinates of the moduli space as well. 

A similar property of holomorphicity is found in two-dimensional
conformal field theories, if one compares the moduli space of the 
four-dimensional theories with the coordinate space $z= x_1 + i x_2$
of the two-dimensional theories. Actually, conformal symmetry
implies that the stress-energy tensor is holomorphic
$\partial / \partial{\overline z}\ T(z) = 0 $.
This relation is stable under analytic
reparametrizations of the coordinate ($z \rightarrow w$), 
\beq 
\frac{\partial}{\partial\overline z} T(z) = 0 \ \ \leftrightarrow \ \ 
\frac{\partial}{\partial\overline w} T(w) = 0 \ ,\qquad
z = \sum_{n \in \Z} \epsilon_n w^{n+1} \ ,
\eeq
and leads to powerful Ward identities which can be solved for the
correlators\cite{cft}. Within this infinite-dimensional covariance of the
theory, the true symmetry transformations are given by the projective
$sl(2,\C)$ subalgebra. To summarize, in conformal field theory
holomorphicity is associated with an infinite-dimensional covariance
and integrability.

In the supersymmetric theories, holomorphicity is similarly stable under
analytic field redefinitions ($\Psi \rightarrow \Sigma$): 
\beq 
\frac{\delta}{\delta\overline\Psi}  {\cal F} ( \Psi) = 0
\ \ \leftrightarrow  \ \ 
\frac{\delta}{\delta\overline\Sigma}  {\cal F} ( \Sigma) = 0 \ ,
\qquad \Psi = \sum_{n\in \Z} \epsilon_n \ \Sigma^{n+1} \ .
\eeq
These non-linear transformations are not familiar in field theory, 
because we usually consider the Ward identities for linear field 
variations\footnote{
In particular, non-linear transformations of $N=2$ chiral
fields are usually disregarded \cite{seiberg}, because they violate 
the Bianchi identity for the photon in the $N=2$ chiral multiplet  
$( D^{i\alpha} D^j_{\alpha} \Psi = 
{\overline D}^i_{\dot\alpha} {\overline D}^{j\dot\alpha} \overline\Psi)$.}.
The true symmetries of the action are restricted to the
finite-dimensional supersymmetric transformations 
$\delta_\xi \Psi = \xi^{i\alpha} Q^i_\alpha \Psi $ which act trivially 
on the moduli, $\delta_\xi a = 0$.

Nevertheless, we believe that there should exist Ward
identities for the infinite-dimensional covariance in field space. 
A necessary condition for this covariance is the possibility
of continuously deforming the moduli space of the
low-energy theory, without changing its main features. 
This is called the isomonodromy property \cite{iso}.
In this paper, we show that the Seiberg-Witten solution is
indeed isomonodromic and discuss some consequent effects.

The Seiberg-Witten solutions of the massive $SU(2)$ theories involve
the holomorphic sections ($a_D(u), a(u)$), whose $(N_f+2)$ 
singularities can be displaced in the moduli space $\{ u \in {\bf C} \}$,
without varying the monodromies around them\footnote{
Another singularity sits at infinity.}. 
Actually, the $N_f$ quark singularities $u
\simeq O ( m^2_i )$ can be moved by varying the masses, and their
monodromies are given by constant integer matrices.
The two additional singularities $ u = \pm O (\Lambda^2 )$ 
also have integer monodromies, but there is only one parameter
for displacing them. 
Therefore, we should allow a further isomonodromic deformation 
of the $u$ variable, which changes the physical moduli space into a
more general, unphysical one. This can be done
because the Seiberg-Witten section is specified by an elliptic curve:
small $u$ deformations correspond to small variations of the
coefficients of the curve which do not generically change the monodromies.
In the particular case of the pure $SU(2)$ theory, 
it is also possible to displace the two singularities by 
$SL(2,{\bf C})$ projective transformations of $u$, which
are invertible in the whole plane and thus leave the monodromies invariant.
We thus conclude that the $SU(2)$ Seiberg-Witten solution is
``covariant'' under $SL(2,\C)$ transformations in the $u$ plane. 

The isomonodromy of the Seiberg-Witten solution has also been 
discussed in the recent Refs.\cite{isopap}\cite{solit}. This is a general
property of integrable systems, which is not too useful in practice:
one actually needs to identify the specific class of integrable systems 
corresponding to the Seiberg-Witten solution, and to characterize it 
by an infinite-dimensional covariance. 
Some interesting relations with integrable systems have been found in 
the Refs.\cite{warner}.

\begin{figure}
\epsfxsize=10cm 
\centerline{\epsfbox{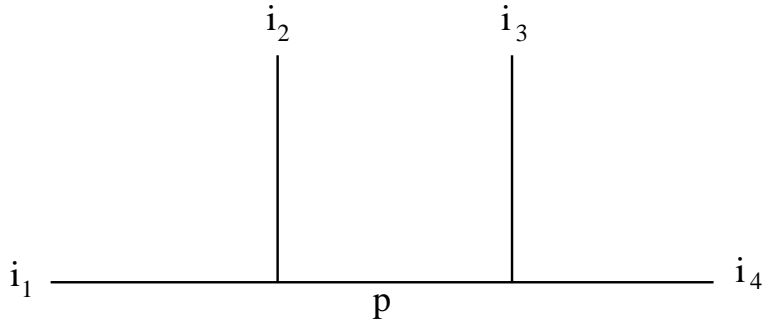}}
\caption{``Duality'' diagram for the $4$-point conformal block.}
\label{fig1}
\end{figure}

The rational conformal field theories are a well-understood class
of isomonodromy problems.
The monodromy properties of the $n$-point conformal blocks 
(holomorphic part of the correlators) have been analysed by Moore
and Seiberg in Ref.\cite{mose}. 
Consider for example the holomorphic part of the four-point function:
\beq 
\langle \phi_1 (z_1)\phi_2(z_2)\phi_3(z_3)\phi_4(z_4) \rangle \propto\  
\sum_{j=1}^q \ a_j \ {\cal F}^{j}_{i_1 i_2 i_3 i_4} 
\left( {(z_1-z_2)(z_3-z_4) \over (z_1-z_3)(z_2-z_4) } \right) \ .
\eeq
In this equation, the $q$ conformal blocks ${\cal F}^j(\eta)$ form
a $q$-dimensional holomorphic section with three branch points at
$\eta=0,1,\infty$, which satisfy a $q$-th order differential equation.
The behaviour of each block ${\cal F}^j$ for $z_1 \to z_2$
is given by the corresponding term in the operator product expansion,
\beq 
\phi_1(z_1) \phi_2(z_2) \simeq \sum_{j=1}^q {(z_1-z_2)}^{h_j-h_1-h_2} 
\phi_j(z_2) \ ,
\eeq
where $h_1,h_2,h_j$ are the conformal dimensions of the fields.
This fusion of fields is described by the ``dual'' diagram in
Fig.(\ref{fig1}).
The monodromy for $z_1$ going around $z_2$ has the diagonal form 
$M_2 = {(\Lambda_2)}_{jj'}= \delta_{jj'} e^{i2\pi (h_j-h_1-h_2)}$;
it is independent of the position of the other singularities, 
because they do not enter in the local operator-product expansion. 
The monodromies around the other singularities are of the form $M_i = U_i
\Lambda_i U^{-1}_i$, where the $U_i$ are the transformations for carrying the
monodromy paths to a common base point. The diagonal matrices $\Lambda_i$
are again determined by the operator-product expansion of the corresponding 
fields; the $U_i$ are the transformations among equivalent sets of 
solutions of the differential equation for ${\cal F}^j$, which
are determined by the universal data of the rational conformal
field theory \cite{dotsenko}. Therefore, all the monodromies 
are independent of the $\{z_i\}$.

Moore and Seiberg have introduced the fusion operator which
acts on the fields and modifies the fusion pattern; namely, 
it replaces the ``$s$-channel'' diagram in Fig.(\ref{fig1}) 
with the ``$t$-channel'' diagram.
Furthermore, there is the braiding operator which exchanges the fields
in the block by analytic continuation in the $z$-plane.
This operator satisfies the Yang-Baxter identity due to the
associativity of exchanges; moreover, the associativity of the
fusion and braiding operators is enforced by the Moore-Seiberg
Pentagonal identity. The Yang-Baxter and Pentagonal identities 
summarize the isomonodromy of the conformal blocks.

In this paper, we show that analogous braiding and fusing operations
can be defined for the Seiberg-Witten sections.
On the other hand, we also find that these holomorphic functions are 
different from the conformal sections, in general. 
In Section $2$, we follow the motion of the singularities 
as the quark masses are varied and find that they braid
and fuse in pairs. 
We define the braiding and fusing operators which
act on the singularities, rather than the ``fields'', and
actually transform their monodromy matrices.
We then show that these operators satisfy the Yang-Baxter
and Pentagonal identities, respectively. 
Actually, these identities express topological properties
which must hold in any isomonodromy problem. 

The physical meaning of these braidings and fusings
in the supersymmetric theory is the following \cite{s-w}: the braiding 
describes the transmutation of a massless quark singularity at weak 
coupling $\left( u\simeq m^2 \ \ {\rm for}\ \ m^2 \gg \Lambda^2 \right)$ 
into a massless monopole one at strong
coupling $ ( u=O(\Lambda^2){\rm \ for \ }  m^2 = O(\Lambda^2) )$. 
The fusing describes the merging of quark and monopole 
singularities into a superconformal singular point \cite{douglas}.
These critical points occur sometimes at finite values
$m_i = O(\Lambda)$ on the trajectory $m_i \rightarrow \infty$ of a 
quark decoupling -- a rather odd renormalization group
behaviour. Actually, we show that the pattern of superconformal points, 
i.e. of the fusings, is determined by the consistency with the braidings; 
namely, it is a physical effect of isomonodromy.

The BPS mass formula for $N=2$ supersymmetric theories with $N_f$ 
quark hypermultiplets of mass $m_f$ is given by \cite{s-w}:
\beq 
m^2 = 2 |Z|^2 \ , \qquad
Z \ = \ n_m a_D (u)+ n_e a(u) + \sum_{f=1}^{N_f} s_f 
\frac{m_f}{\sqrt{2}} \ ,
\label{bps}\eeq
where $(n_m,n_e,s_f)$ are the magnetic, electric and (pseudo)-baryonic
quantum numbers, respectively. 
The monodromy transformations of $(a_D,a)$ contain additive terms
proportional to the masses, which were not completely understood
in the literature. 
In Section $2.3$, we use the Yang-Baxter and Pentagonal identities 
as equations to determine completely these affine terms;
as an input, we use the matching between the $N_f$ and the $(N_f-1)$ 
theories at the quark decouplings.
This algebraic derivation is simpler than the direct analytic continuation of 
the general Seiberg-Witten sections.
The complete monodromy matrices of the $N_f=1$ theory are derived 
in the text, while those of the $N_f=2$ and $3$ theories are found in 
the Appendix A. Moreover, simple explicit expressions for the Seiberg-Witten 
sections are obtained in Appendix B for the cases of maximally fused 
singularities; their monodromy transformations are then used to 
check the results by the algebraic approach.

Next, we use the complete monodromies to determine the full 
spectrum of the pseudo-baryonic quantum numbers.
The $(n_m,n_e,s_f)$  quantum numbers in (\ref{bps}) are also transformed 
under the monodromy of $(a_D,a)$, so that $Z$ remains invariant; 
this is the spectral flow, which should map the spectrum 
into itself. Following Ref.\cite{bf1}, we use this flow as 
a self-consistent constraint which determines the complete 
weak-coupling spectrum.

In Section $3$, we study more closely the analytic properties of the
Seiberg-Witten section in the first non-trivial case of three
singularities ($N_f=1$). We rewrite the Picard-Fuchs
equations satisfied by the section \cite{yang} 
into a Fuchsian second-order 
differential equation, which is the traditional setting for 
studying isomonodromy \cite{yoshida}. 
This property implies a set of differential integrability conditions 
for the residues of the single-pole terms (the accessory parameters)
and the apparent singularity in the Fuchsian equation.
For three singularities, these conditions reduces to a single non-linear
differential equation, the Painlev\'e VI equation for the apparent
singularity. We find that this is rather remarkably satisfied by the
Seiberg-Witten solution: this yields an explicit proof of
isomonodromy, which confirms the previous, intuitive arguments. 
As an outcome of this analysis, we 
find a new first-order differential equation for the mass variation
$(\partial a_D/\partial m,\partial a/\partial m)$. 

In Section $4$, we directly identify some Seiberg-Witten sections with 
the conformal blocks of suitable conformal field theories. 
The two-dimensional monodromies imply the operator product expansion
$\phi_1 \cdot \phi_2 = \phi + \phi^\prime$. This is reproduced,
for example, by the primary field
$\phi_{1,2}$ of the minimal conformal models with 
central charge $c(p,p') = 1 - 6 {(p-p')}^2/{pp'}$.
The presence of logarithmic singularities requires that the
dimensions of the fields $\phi$ and $\phi'$ are equal,
$h = h^\prime$; this occurs in the so-called logarithmic conformal 
field theories with $c=c(1,p')$ \cite{gurarie}. We consider 
the simplest Seiberg-Witten sections with three singularities, 
which describe the $N_f=0$ theory and $N_f=1,2,3$ ones 
with $(N_f+1)$ singularities fused into a superconformal point.
Indeed, we are able to identify these sections with the four-point 
conformal blocks of the theories with $c=c(1,2)$, $c(1,3)$, $c(1,4)$ 
and $c(1,6)$, respectively.

This correspondence is not sufficient to establish
the complete equivalence of the supersymmetric and conformal 
isomonodromy problems.
Actually, the previous three-singularity sections
are necessarily Hypergeometric functions, 
which are conformally covariant under the $SL(2,{\bf C})$ projective
transformations of the $u$-plane, and thus can be
represented by (quasi)-primary conformal fields \cite{cft}.
On the contrary, the general supersymmetric sections with
$(3+N_f)$ singularities are shown to be non-covariant under $SL(2,{\bf C})$,
and thus cannot correspond to conformal blocks.
It remains an open question whether the 
supersymmetric isomonodromy problem can be characterized by an
infinite-dimensional covariance.
In Appendix C, we recall some general relations among isomonodromy,
integrability and conformal symmetry, which complement the analysis
of Section $4$.

In conclusion, we believe that it is interesting to investigate 
the relations of the Seiberg-Witten theory with other
isomonodromy problems which are not conformal in general,
like the topological field theories \cite{topft}, the matrix models 
and quantum Liouville theory \cite{df}. Actually, a relation with the
topological theories has already been suggested in the Refs.\cite{moro}
\cite{yang2}: starting from the supersymmetric effective action, these
Authors found an analogue of the fusion rules of the 
topological fields, which satisfy an associativity condition
similar to the one discussed here and encode common isomonodromic properties.


\section{ Braiding and Fusing of Monodromies and Their Identities}

\begin{figure}
\epsfxsize=14cm 
\centerline{\epsfbox{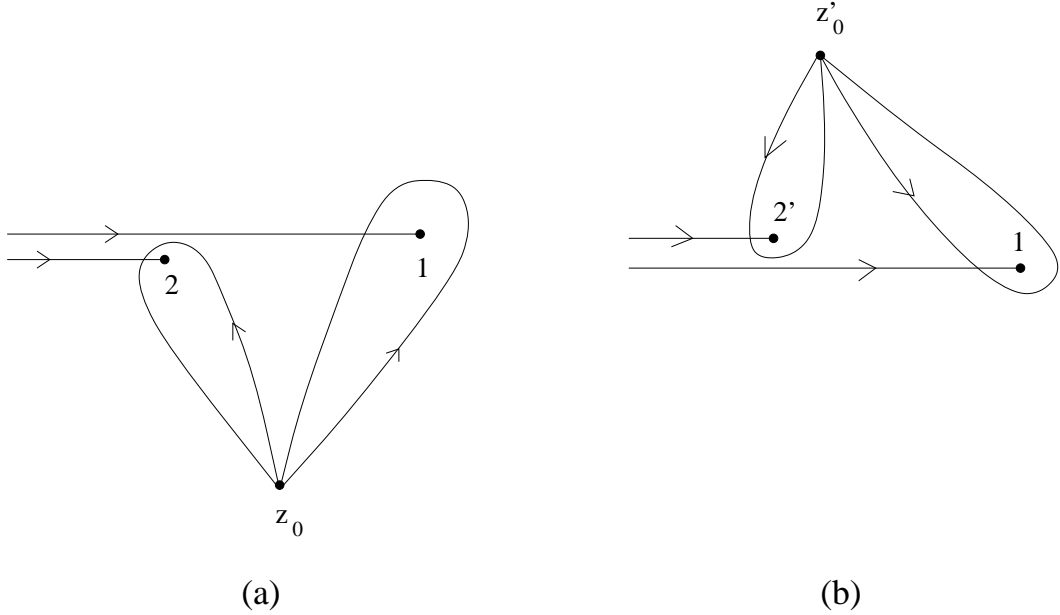}}
\caption{Description of the monodromies for the $N_f=0$ theory.}
\label{fig2}
\end{figure}


\subsection{ Representation of the Monodromy Group}

We begin by introducing our notation for describing the monodromies. 
Let us first consider the  $N_f = 0$, $SU(2)$ theory for simplicity,
whose section
$(a_D(u),a(u))$  has two singularities located at $u_1 = \Lambda_0$ and
$u_2 = - \Lambda_0$.
We can represent the section by a two-dimensional vector $ {\bf a} =
(a_D,a)$ and the monodromy transformation  by matrix multiplication:
\beq
\left( \begin{array}{c}  a_D \\ a \end{array} \right) ( e^{2 \pi i}(u
- u_i) + u_i) = M_i 
\left( \begin {array}{c} a_D \\ a \end{array}\right) \ ,\ \qquad 
i = 1,2\ .
\label{modmat}\eeq
The corresponding transformation of the vector of magnetic and electric
charges ${\bf n} = (n_m, n_e)$ is by right multiplication 
${\bf n} \rightarrow{\bf n} M_i^{-1}$,
such that the BPS mass formula $ Z = {\bf n  \cdot a}$ in Eq.(\ref{bps}) 
remains invariant.
We recall from Ref.\cite{bf1} the monodromy matrices \footnote{
In the normalization valid for $N_f > 0$ \cite{s-w}.}:
\beq
M_1 = \left( \begin{array}{cc}  1 & 0 \\ -1 & 1 \end{array} \right)\ ,\  
M_2 = \left(\begin{array}{cc} 3 & 4 \\ -1 & -1 \end{array} \right)\ ,\ 
M_{2'}=\left(\begin{array}{cc} -1 & 4 \\ -1 & 3 \end{array} \right)\ ,\ 
\label{mod0}\eeq
which satisfy the relations,
\barr
M_1 M_{2'} = M_2 M_1 = M_{\infty}\ ,\qquad
M_{\infty}=\left(\begin{array}{cc} -1 & 4 \\ 0 & -1
\end{array}\right) \equiv M_\infty^{(0)}\ .
\label{m2m2p}
\earr
These non-Abelian monodromies depend on the position of the
base point for the loop and on the location of the cuts. 
As explained in Ref.\cite{bf1}, the analytic continuation of the section
$(a_D,a)$ depends on the Riemann sheet of 
$u\sim u_i$ in Eq.(\ref{modmat}): for example, there are
two possibilities $M_2$ and $M_{2'}$ for the second singularity,
due to presence of the cut emanating from the first 
singularity. In principle, one should keep track of 
the Riemann sheets when composing the monodromies.

Our representation of the monodromies is shown in Fig.(\ref{fig2}):
the two elementary loops of Fig.(\ref{fig2}a) can be associated to 
the monodromies $M_1\equiv M_1(z_0)$ and $M_2\equiv M_2(z_0)$, 
because they reproduce $M_{\infty} = M_2 M_1$ by composition; 
similarly, the monodromies $M_1$ and $M_{2'}\equiv M_2(z_0')$ 
correspond to the loops in Fig.(\ref{fig2}b). 
Our graphical rules do not faithfully reproduce all the analytic
properties of the Seiberg-Witten section, but will be sufficient to 
describe the monodromy group and the braiding and fusing operators; 
they are inspired by the previous analysis of Chern-Simons 
monodromies in a ``singular gauge'' \cite{ccv}:
\begin{itemize}
\item 
We use a single plane with cuts, which corresponds to an initial,
conventional choice of the Riemann sheet, i.e. of the patch 
and its monodromies;
\item 
We can imagine that all the non-trivial monodromy is concentrated
at the crossings of the path with the cuts: this could be obtained by
an analytic reparametrization of the $u$ coordinate, which leaves 
the monodromies invariant \cite{ccv};
\item
The monodromies are invariant if the base point is moved in the
plane without crossing a cut; 
\item
The monodromies with fixed base point are invariant under 
the motion of the singularities: the deformed paths following 
them can cross a cut but not a singularity.
\end{itemize}

\begin{figure}
\epsfxsize=14cm
\centerline{\epsfbox{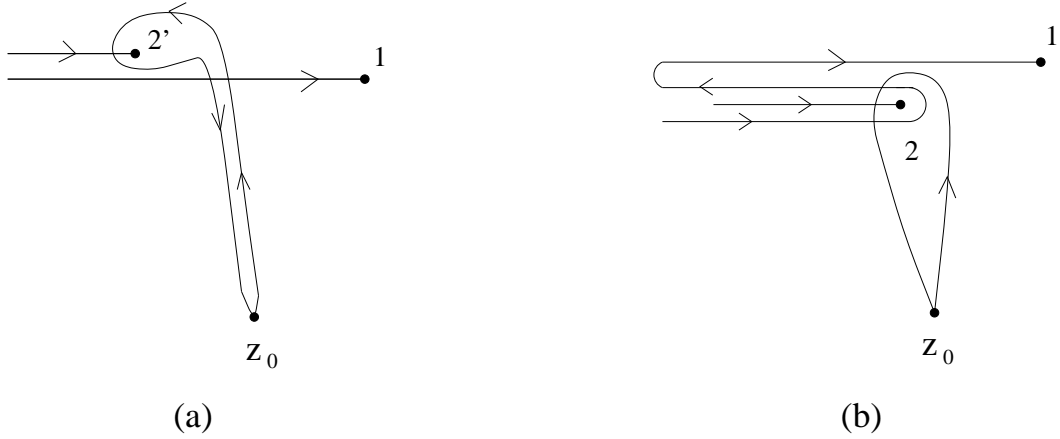}}
\caption{Changes of patch}
\label{fig3}
\end{figure}

In Fig.\ref{fig3}, we use these rules to deal with: $(a)$, the change
of base point; $(b)$, the motion of the singularity $(2)$ through the cut
of $(1)$. In order to compare $M_2$ with $M_{2'}$, we imagine to move $(2)$
in Fig.(\ref{fig2}a) above the cut of $(1)$: since $M_2(z_0)$ is invariant,
we find that
\beq
M_2(z_0) = M_1 M_{2'} M_1^{-1}\ \longrightarrow\  
M_{2'} = M_1^{-1} M_{2} M_1 \ ,
\label{firstyb}\eeq
which is in agreement with (\ref{m2m2p}).
Fig.(\ref{fig3}b) describes the opposite motion of $(2')$ going
above the cut of $(1)$, and shows that the singularity $(2)$
can be obtained by ``dressing'' $(2')$, which amounts to 
the conjugation (\ref{firstyb}) again.
Finally, we choose to send all the cuts to infinity
along the same direction; in this case, $M_{\infty}$ is 
independent of the base point.


\subsection{Braiding and Fusing Operators and ``Duality'' Identities}

Let us first consider the $N_f=1$, $SU(2)$ theory as an example;
the properties of the Seiberg-Witten solution 
have been analysed in Refs.\cite{douglas}\cite{bf1}, 
which contains some useful background. The elliptic curve is:
\beq
y^2 = x^3 - u x^2 + m\frac{\Lambda^3_1}{4} x - \frac{\Lambda^6_1}{64} \ .
\label{el}\eeq
This determines the Seiberg-Witten section ($a_D(u), a(u)$) by 
$u$-integration of the Abelian integrals,
\beq
\frac{da}{du} = \frac{\sqrt{2}}{8\pi} \oint_{\gamma_1} \frac{dx}{y} \ ,
\qquad\ \frac{da_D}{du} = \frac{\sqrt{2}}{8\pi} \oint_{\gamma_2}  
\frac{dx}{y} \ ,
\label{perin}\eeq
and by matching with the known asymptotics. The singularities 
in the $u$-plane occur when one period $\gamma_i$ vanishes and
are given by the zeroes of the discriminant of the curve\footnote{
Hereafter, we shall often use the normalization $\Lambda^{4-N_f}_{N_f} = 8$.}
(\ref{el}):
\beq
\Delta (u) \ = \ 8 \left(u^3 - u^2 m^2 - 9 u m + 8 m^3 + 
\frac{27}{4}\right) \ . 
\label{delta}\eeq
\begin{figure}
\vspace{-2.cm}
\epsfxsize=10cm \centerline{\epsfbox{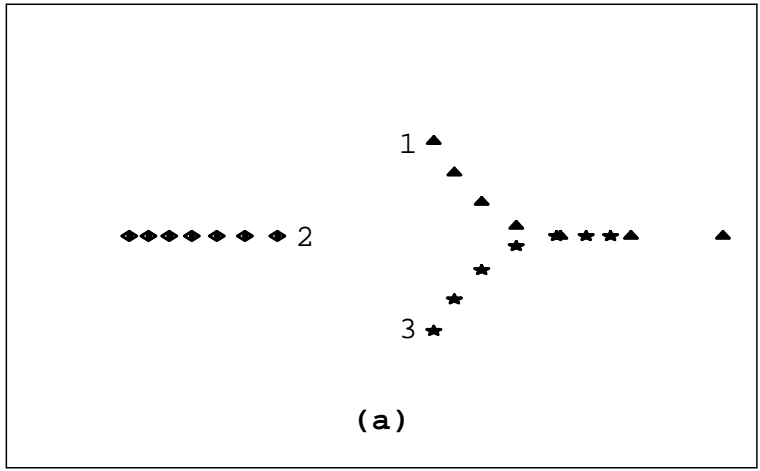}}\vspace{-3.5cm}
\epsfxsize=10cm \centerline{\epsfbox{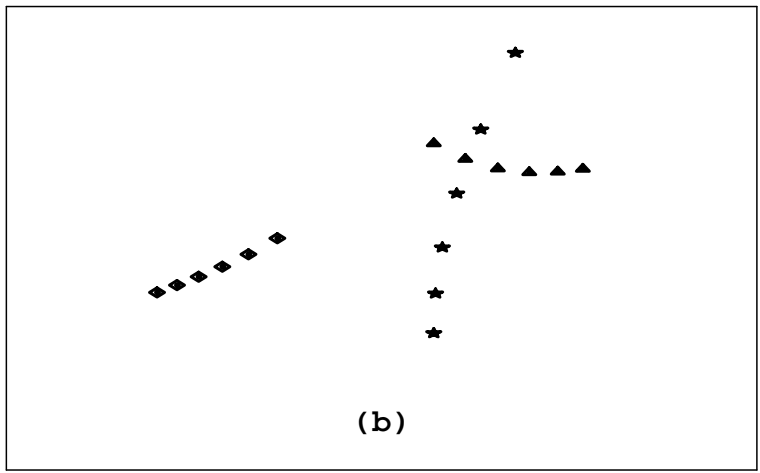}}\vspace{-3.5cm}
\epsfxsize=10cm \centerline{\epsfbox{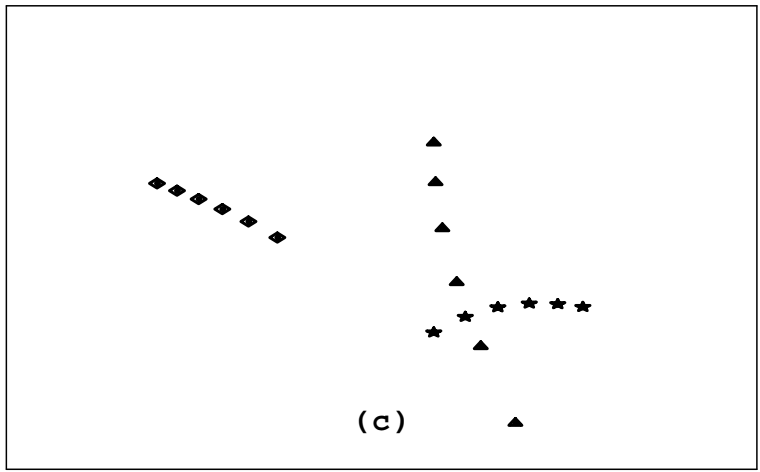}}\vspace{-2.cm}
\caption{Motion of the three $N_f=1$ singularities in the $u$-plane:
$(a)$, $Arg(m)=0$; $(b)$, $Arg(m)>0$; and $(c)$, $Arg(m)<0$.}
\label{fig4}
\end{figure}
The singularities for $m=0$ are located at 
$u = ( u_0 e^{i\pi/3}, - u_0, u_0 e^{-i\pi/3} )$, with $u_0=3/4^{1/3} $. 
When the mass is
switched on, their motion in the $u$-plane is shown in Fig.(\ref{fig4}):
for real positive mass, (case (a)), the singularities $(1)$ and $(3)$
fuse together at $u=u_c=3$ and $m=m_c=3/2$, 
and then split again and move on the real axis. 
The ($u_c,m_c)$ point describes a superconformal theory, which has
been  discussed in the Refs.\cite{douglas}. 
After the split, one singularity approaches the value $u
\simeq m^2$ for large $m$, which corresponds to the weak-coupling
quark singularity. 
The other two singularities asymptotically become those of the
$N_f=0$ theory: actually, after matching the $\Lambda$-parameters by
$\Lambda_0^4 = m \Lambda^3_1$, their position tends to
$u= \pm \Lambda^2_0$.
This motion of singularities satisfies the expected
``renormalization-group flow''\footnote{
Actually, this is the explicit tuning of a parameter.}
 from the $N_f$ theory to the
$( N_f-1 )$ theory when one quark becomes very massive and decouples.
Note, however, the rather unusual presence of a critical point at 
finite bare mass, whose origin will be explained later. 

Fig.(\ref{fig4}b) shows the motion of the singularities for $m$ growing
along the ray $Arg(m) = \pi / 6$ ( Fig. 1 (c) is the case 
$Arg(m) = - \pi/6$ ). Here one sees that the
trajectories of the singularities $(1)$ and $(3)$ cross each other 
before approaching the decoupling  asymptotic configurations 
$u= \pm \sqrt{m \Lambda^3_1}$ and $u = m^2$, on the asymptotic rays 
$Arg (u) = Arg (m) /2$ and $ Arg (u) = 2 Arg (m)$, respectively.  
These trajectories corresponds to an exchange of the
singularities, which pass by the crossing point at different $m$
values. One can see that the singularities braid clockwise or
counter-clockwise for $Arg (m) >0$ and $Arg (m) <0$, respectively.
Note also that the braiding and fusing of other pairs of singularities 
similarly occurs for $\pi/3 < Arg (m) < \pi$ and $-\pi < Arg (m) <
-\pi/3$, due to the ${\Z}_3$ symmetry of the $m=0$ theory.

Similar braidings and fusings of the singularities of an
holomorphic section have been analysed in conformal field 
theory by Moore and Seiberg \cite{mose}. 
They were able to define the operators 
which specify the analytic continuation 
and the fusion of the fields in the conformal blocks, respectively. 
Here, we do not have a representation of the Seiberg-Witten 
section in terms of ``fields'' (see however 
Section $4$); therefore, we shall describe the simpler
braiding and fusing operators acting on 
the monodromy matrices themselves.

In the theories with massive quarks, the monodromy transformations 
acquire an additive term proportional to the mass \cite{s-w};
for the sake of simplicity, we shall first discuss the monodromies of
$(d a_D/d u, d a/d u)$, which do not have this term.
Let us also recall from Ref.\cite{bf1}  the general formula
for the monodromy around the singularity caused by $k$ BPS states 
with charge ${\bf n} = (n_m,n_e)$ becoming massless:
\beq
M_{k (n_m,n_e)} = \left( \begin{array}{cc}  1 + n_e n_m k & n_e^2 k  \\
-n_m^2 k & 1 - n_e n_m k \end{array} \right) .
\label{mgen}\eeq
These are the ``elementary'' monodromies, made by the analytic continuation
along a small loop near the singularity. 
The following properties:
\beq
M_{k(n_m,n_e)} = M_{k(-n_m,-n_e)} =M_{k/\ell^2(\ell n_m,\ell n_e)} \ ,
\qquad  M^{-1}_{k(n_m,n_e)} =M_{-k(n_m,n_e)} \ ,
\label{mprop}\eeq
imply a one-to-one correspondence between monodromy matrices
and charge vectors ${\bf n}$ for $n_m \le 2$, as in the present theories.
We also need the general formula for the conjugation of monodromies:
\barr
M_{(n_m,n_e)} &\to& 
M^{-1}_{k(n_m',n_e')}\ M_{(n_m,n_e)}\ M_{k(n_m',n_e')} \ \ ,\nl
(n_m,n_e) &\to & (n_m,n_e) + (n_m',n_e') k\left[n_m n_e' -n_e n_m' \right]\ .
\earr
This shows that the monodromies of ``mutually local'' particles
$( {\bf n}\times{\bf n}'=0)$ commute among themselves.

\begin{figure}[hh]
\epsfxsize=14cm \centerline{\epsfbox{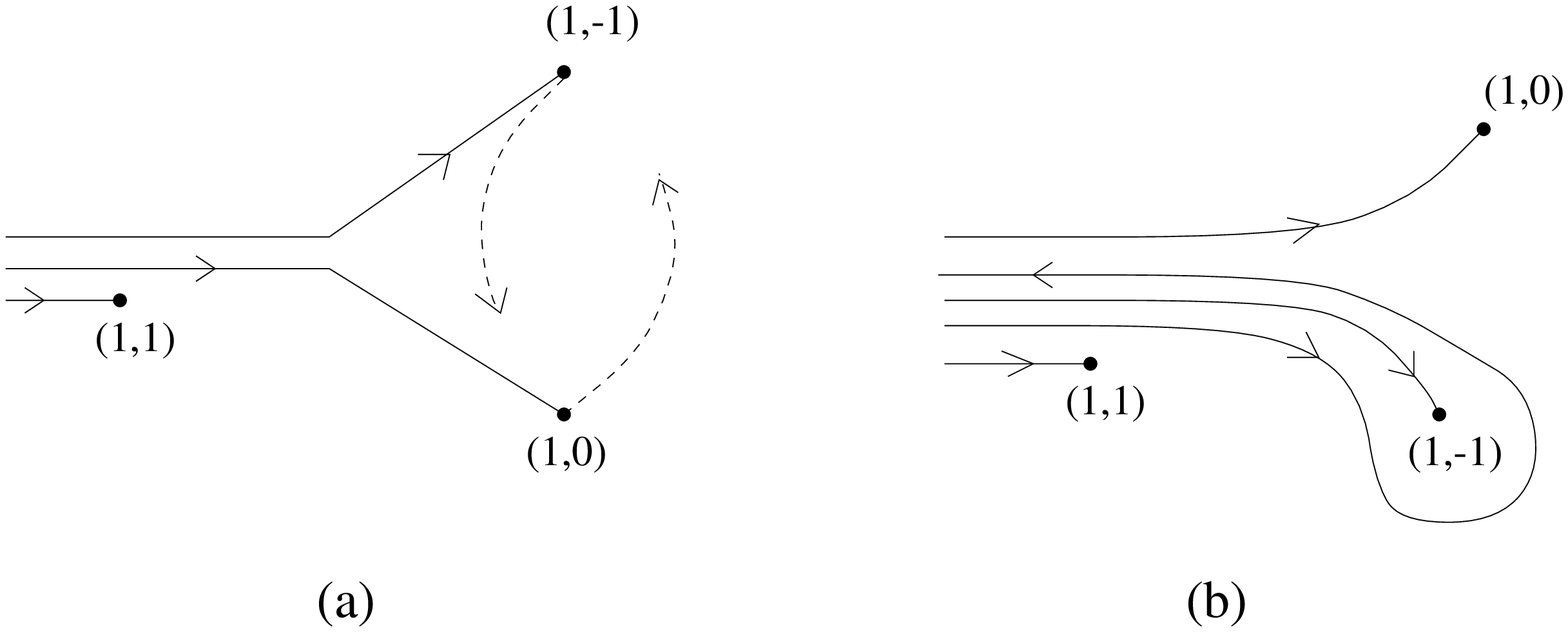}}
\epsfxsize=14cm \centerline{\epsfbox{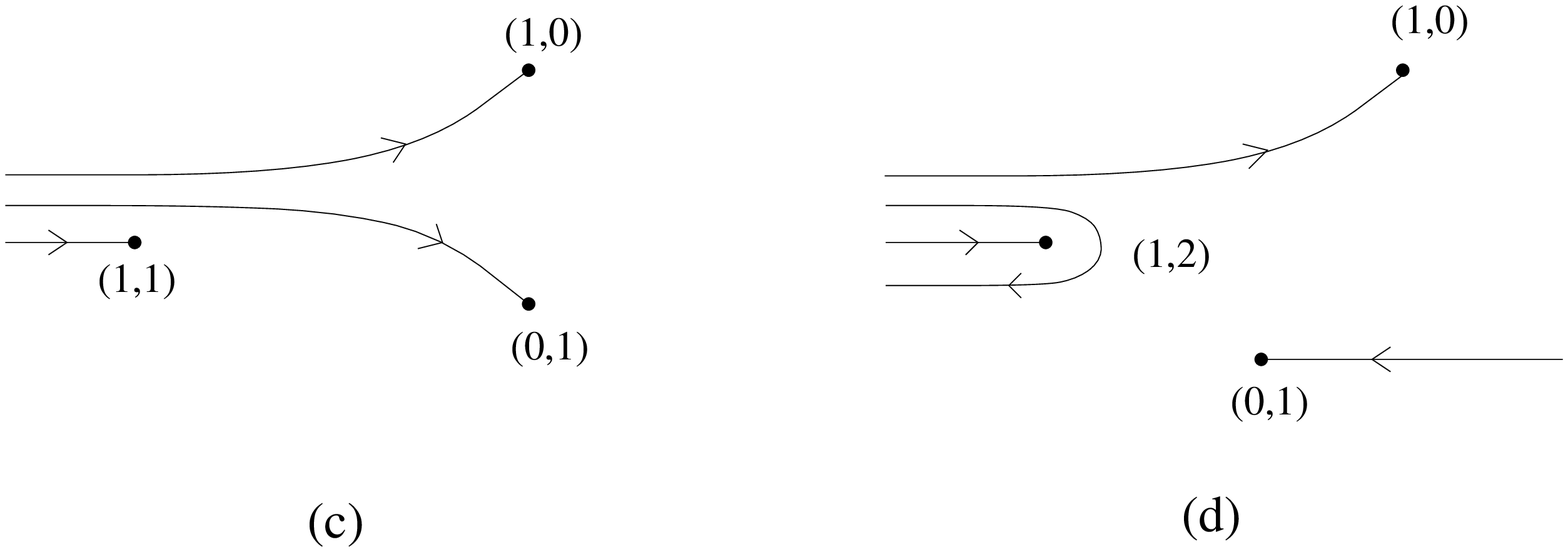}}
\caption{Deformations of cuts associated to the singularity exchange and
the quark decoupling.}
\label{fig5}
\end{figure}

A convenient choice of patch for the $N_f = 1$ section is
shown in Fig.(\ref{fig5}a), following our previous conventions.
The elementary monodromies around the three singularities can be
carried to a common base point located in the bottom right corner 
of the figure; their values are \cite{s-w}:
\beq
M_1=M_{(1,-1)}\ , \quad M_2=M_{(1,1)}\ ,\quad M_3=M_{(1,0)}\ . 
\label{m1m2m3}\eeq
The composition giving the monodromy at infinity is,
\beq
M_2 M_3 M_1 = M^{(1)}_{\infty}=\left(
\begin{array}{cc} -1 & 3 \\ 0 & -1 \end{array} 
\right)\ , 
\label{minf1}\eeq
as it can be easily recognized from the order of the cuts at infinity.
Let us recall that the values of the electric charge are patch dependent,
due to the phenomenon of ``democracy of dyons''\footnote{
See Ref.\cite{ken} for a complete discussion.}: 
any other choice of three consecutive values in Eq.(\ref{m1m2m3})
would be equivalent by rotating the $\theta$-angle, i.e.
$Arg(u)$ at infinity.
It is important to keep this arbitrariness in mind when comparing
different results in the literature.

Let us now describe the transformation of the three monodromies 
(\ref{m1m2m3}) corresponding to the counter-clockwise exchange of 
two singularities in (Fig.(\ref{fig4}c))  \\ $\ (Arg(m)<0)$.
As shown in Fig.(\ref{fig5}a), the singularity $(1,-1)$ 
crosses the cut of $(1,0)$ and goes into another Riemann sheet.
This can be taken into account by deforming the cut and dressing the 
singularity according to Fig.(\ref{fig5}b): 
the result is a new elementary monodromy around this singularity,
Fig.(\ref{fig5}c), as seen in the original Riemann sheet, and 
from the same base point. This is obtained by the conjugation:
\beq
M_{(1,-1)} \to M_{(1,0)} M_{(1,-1)} M_{(1,0)}^{-1}=M_{(0,1)}\ .
\label{trans}\eeq

Therefore, the massless dyon has become a massless quark.
Next, Fig.(\ref{fig5}d) shows the configuration which should be
attained for the decoupling of this massive quark: its cut
should go to infinity as well, such that the moduli space at finite $u\ $
$(\Lambda_0\sim u\ll m)$ only contains the singularities and the cuts
of the $N_f=0$ theory in Eq.(\ref{mod0}) and Fig.(\ref{fig2}a). 
This requires a further conjugation of the singularity $(2)$,
yielding ${\bf n}_2 =(1,1) \to (1,2)$.
In conclusion, the quark decoupling is characterized by the following 
relation between the singularities at infinity in the two theories:
\beq
M^{(1)}_\infty = M_{(0,1)}\ M_{(1,2)}\ M_{(1,0)}=
M_{\rm quark}\ M^{(0)}_\infty \ .
\eeq 

The two characterizations (\ref{trans}) of the same singularity as a
dyon and a quark pertain to different Riemann sheets, and there is 
no discontinuity in the physics. 
Within each patch, it is possible to vary smoothly $u$ from a 
given point (e.g. the base point), reach any singularity and 
find that a certain BPS state is massless there.
The discontinuity is found when comparing 
the two asymptotic behaviours, $m\sim\Lambda_1$ 
(in the strong coupling region) and $m\gg\Lambda_1$ (in the
weak coupling region): these are necessarily
described by two different Riemann sheets due to 
the topology of the motion of the singularities and the cuts.
Actually, if we stay always in the same patch, we find that the cuts 
get tangled and, asymptotically, it becomes
impossible to reach smoothly all the singularities from a common
point. Therefore, the motion of the 
singularities implies an asymptotic change in the charges of the 
massless particles, due to the necessary changes of patches;
this is a rather remarkable consequence of the physics being
described by a multi-valued holomorphic function.
Moreover, upon breaking the $N=2$ supersymmetry to $N=1$ \cite{s-w},
this property verifies the equivalence of the confinement phase
(monopole condensation) and the Higgs phase (quark condensation) 
in gauge theories with quarks in the fundamental representation 
\cite{peskin}.

\begin{figure}
\epsfxsize=14cm \centerline{\epsfbox{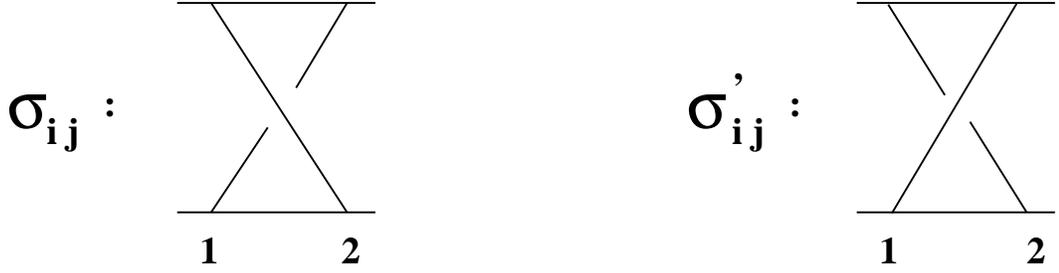}}
\caption{Braid operators $\sigma_{ij}$ and $\sigma_{ij}'$ .}
\label{fig60}
\end{figure}

\begin{figure}
\epsfxsize = 10 cm 
\centerline{\epsfbox{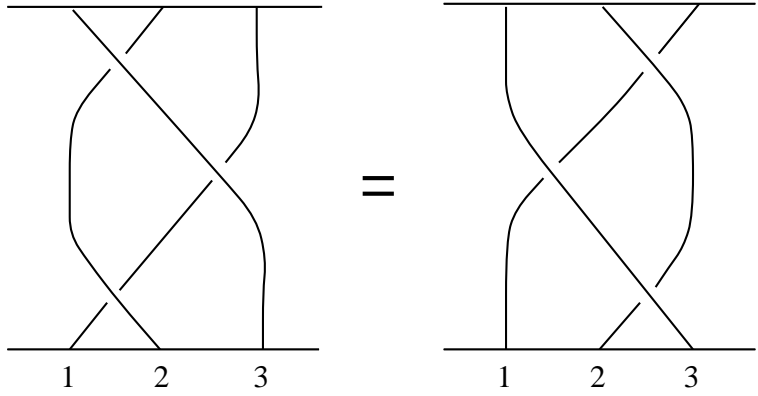}}
\caption{The Yang-Baxter identity.}
\label{fig6}
\end{figure}

The counter-clockwise exchange of two singularities in Fig.(\ref{fig5}a,b)
can be associated with one of the braid operators $\sigma_{ij}$, 
$i,j=1,2,3$, which act in the tensor product 
$V_1 \otimes V_2 \otimes V_3$ of vector spaces of the matrices $M_i \in V_i$:
\beq
\sigma_{1 3} \left( 
\begin{array}{c} M_1 \\ M_2 \\ M_3 \end{array}
\right) = \left( 
\begin{array}{c} M_3 M_1 M_3^{-1} \\ M_2 \\ M_3 \end{array}
\right)\ .
\eeq
The clockwise exchange occuring for $Arg (m)>0$ (Fig.(\ref{fig4}b))  
yields another operator: 
\beq
\sigma_{1 3}' = \left( 
\begin{array}{c} M_1 \\ M_2 \\ M_3 \end{array}
\right) = \left(
\begin{array}{c} M_1 \\ M_2 \\ M_1^{-1} M_3 M_1 \end{array}
\right)\ .
\eeq
These two inequivalent operators are depicted in Fig.(\ref{fig60}), 
in the standard notation for braids, and are
actually related by $\sigma_{i j}' = (\sigma_{j i})^{-1}$
(Note that the indices are associated to the threads in our notation).
The $\sigma_{i j}$ yield a representation of the braid group 
${ \cal  B}_3$ in the space of monodromy matrices, and
satisfy the Yang-Baxter identity,
\beq
\sigma_{2 3}\ \sigma_{1 3}\ \sigma_{1 2} = 
\sigma_{1 2}\ \sigma_{1 3}\ \sigma_{2 3} \ ,
\eeq
which states the associativity of braids, as shown in Fig.(\ref{fig6}).
The required one-dimensional ordering of monodromies can be recognized
from the position of the cuts as they cross the $u\to\infty$ circle. 

\begin{figure}
\epsfxsize = 9.5 cm 
\centerline{\epsfbox{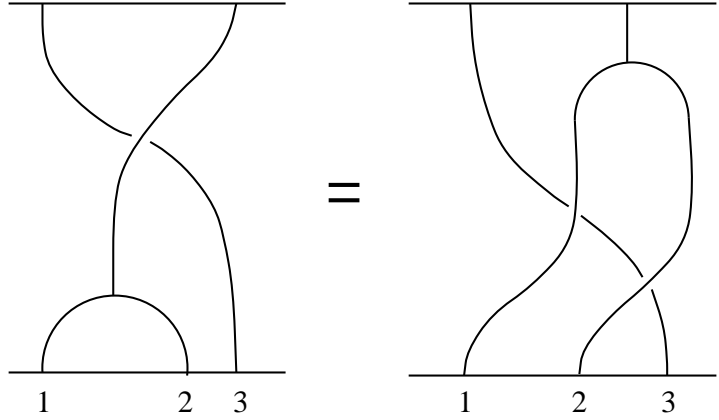}}
\caption{The Pentagonal identity.}
\label{fig7}
\end{figure}

The fusion of two singularities, occurring for $Arg(m) = 0$, 
Fig.(\ref{fig4}a),
can be similarly associated to the fusion operator, which is a
mapping $f_{31} : V_1 \otimes V_2 \otimes V_3 \longrightarrow 
V_{31} \otimes V_2$:
\beq
f_{31} \left( \begin{array}{c} M_1 \\ M_2 \\ M_3 \end{array} \right) = 
\left( \begin{array}{c} M_3 M_1 \\ M_2 \end{array}\right) \ ,
\label{fuse}\eeq
where the ordering in the product $M_c=M_3M_1$ is determined by
the invariance of $M_\infty$.
An analogous fusion operation has been defined 
in conformal field theory \cite{mose} and has been shown to satisfy an
associativity condition with the braids, known as the Pentagonal
identity. In our context, this reads:
\beq
\sigma_{2 3}\ f_{1 2} = f_{1 2}\ \sigma_{1 3}\ \sigma_{2 3} \ ,
\eeq
and is depicted in Fig.(\ref{fig7}). Actually, this is satisfied
by our operators $\sigma_{ij}$ and $f_{ij}$.
The fusion (\ref{fuse}) produces a new type of
singularity, whose monodromy does not belong to the BPS family (\ref{mgen}):
it satisfies $M_c^3 = -1$ and thus corresponds to a power law
singularity $u^\al$ with exponents $ \alpha = \pm 1/6$. 
This scale invariant behaviour is characteristic of a superconformal 
theory, which occurs when two or more mutually non-local particles 
become simultaneously massless \cite{douglas}. 
After fusing, the singularities
split again into a quark and a monopole ones following the usual
decoupling pattern,  $M_c = M_{(0,1)}M_{(1,0)}$.
In this case, there is a true discontinuity in passing from
the patch $(m\sim\Lambda)$ to the asymptotic one $(m\to\infty)$.
Let us remark that the fusion is a necessary 
consequence of the discontinuity of braiding when the mass is
varied continuously from $Arg(m) > 0$ to $Arg(m) < 0$.
This mathematical consistency implies the presence of the strange
critical point in the moduli space along the real trajectory 
$0<m<\infty$; therefore, this is a peculiar physical consequence of 
isomonodromy.

The corresponding analysis of braiding and fusing in
the $N_f = 2,3$ theories is described in Appendix A: 
it confirms the properties of the braid operators discussed here,
and describes the interesting patterns of 
quark decouplings with the associated superconformal points.


\subsection{Complete Monodromies and Pseudo-Baryonic Quantum Numbers}

The complete monodromies of $(a_D,a)$ contain the additive term
$\sum_{f=1}^{N_f}\ (q_{Df},q_{f})\ \ m_f/\sqrt{2}$, with
$q_{Df},q_{f}\in {\bf Z}/2$, 
which is due to the poles occurring in the Abelian differential 
after their integration in $u$ \cite{s-w}. These affine transformations 
can be represented in matrix notation as follows:
\beq
\left( \begin{array}{c} a_D \\ a \\ m_f/\sqrt{2}\end{array} \right)
\longrightarrow {\cal M}
\left( \begin{array}{c} a_D \\ a \\ m_f/\sqrt{2}\end{array}\right)
=\left(  \begin{array}{c|l}
M & {\disp q_{Df} \atop\disp q_f} \\ \hline {\bf 0} & {\bf 1}
\end{array} \right)
\left( \begin{array}{c} a_D \\ a \\ m_f/\sqrt{2} \end{array}\right) \ .
\label{com} \eeq
The matrix ${\cal M}$ acts on the complete vector of quantum
numbers ${\bf n}  = (n_m,n_e,s_f)$ by ${\bf n} \longrightarrow {\bf n}
{\cal M}^{-1}$, as before. As discussed by the recent literature 
\cite{ferrari}\cite{ken}, the pseudo-baryonic charges $\{s_f\}$
differ from the true baryonic charges $\{S_f\}$, 
due to the mixing of the explicit mass terms in the BPS formula
with analogous pieces coming from $a_D(u,m_f)$. 
The baryonic charges have been computed by independent 
semiclassical methods in Ref.\cite{ken}, while the pseudo-baryonic 
numbers are important missing data for the mass spectrum.
In order to control these possible shifts by mass terms,
the normalization of $a_D(u,m_f)$ should be better 
specified, such that the $\{s_f\}$ become unambiguous: 
we adopt the following decoupling condition,
\beq
\lim_{m_i\to\infty} {\bf a}^{(N_f)} \left(u;m_1,\dots,m_i,\dots \right) = 
     {\bf a}^{(N_f-1)}\left(u;m_1,\dots,m_i\!\!\!\!\! / \ ,\dots\right)\ ,
\label{decap}\eeq
with $i\in(1,\dots,f)$; this gives a recursive definition
of ${\bf a}^{(N_f)}=(a_D,a)^{(N_f)}$ in terms of the well-defined 
${\bf a}^{(0)}(u)$. Note that
this limits also implies the matching of patches between the two
theories at quark decoupling, as in Fig.({\ref{fig5}d). 

The determination of the complete monodromies (\ref{com}) requires 
the explicit expression of the Seiberg-Witten section
\cite{lag}\cite{brand}\cite{bf2}, 
whose analytic continuation is rather difficult, in general. 
It follows that these monodromies, as well as the $\{s_f\}$ numbers, 
have not been completely found in the literature 
\cite{brand2}\cite{bf2}\footnote{
Note, however, that it is possible to identify $\{s_f\}\equiv\{S_f\}$
by choosing a boundary condition for $a_D^{(N_f)}$ different
from (\ref{decap}), and then use the independent semiclassical determination
of the $\{S_f\}$ \cite{ken}.}.
In the following, we shall determine the complete $N_f=1$ monodromies
by using the braiding and fusing identities as equations. 
The corresponding analysis for $N_f=2,3$ is given in the Appendix A.
The inputs for these equations are:
\begin{itemize}
\item 
The known $SL(2,{\bf Z})$ monodromies (\ref{m1m2m3}).
\item 
The full monodromy of the quark singularity 
$a \sim - \epsilon\ m/\sqrt{2}$ , $\epsilon = \pm 1$, which can be deduced
at weak coupling from the one-loop calculation \cite{s-w}:
\beq
{\cal M}_{(0,1,\epsilon)} = \left(
\begin{array}{ccc}
1 & 1 & \epsilon \\
0 & 1 & 0 \\
0 & 0 & 1 \end{array} \right)\ ,\qquad \epsilon=\pm 1\ .
\eeq
Note that the sign ambiguity  $\epsilon$ is odd under the parity 
$P : a \to -a$; it is a residual gauge invariance in the
moduli space which will remain undetermined: it corresponds 
to the choice of the square-root branch cut for $a\sim\pm\sqrt{u}\to\infty$.
\item 
The matching of the $N_f$ monodromies with the $(N_f-1)$ ones 
at quark decoupling; in particular the affine terms vanish for $N_f=0$.
\end{itemize}

We write the full monodromies at strong coupling, Fig.(\ref{fig5}a), as
${\cal M}_{(1,-1,r)}$, ${\cal M}_{(1,1,s)}$, ${\cal M}_{(1,0,t)}$, 
leaving their affine terms unknown, and plug them into the 
previous braid relations, Eq.(\ref{trans}) and Fig.(\ref{fig5}d):
\barr
{\cal M}_{(1,0,t)} {\cal M}_{(1,-1,r)} {\cal M}^{-1}_{(1,0,t)} & =&
{\cal M}_{(0,1,\epsilon)} \qquad ({\rm quark \  monodromy})\ ,\nl
{\cal M}_{(0,1,\epsilon)}^{-1} {\cal M}_{(1,1,s)} 
{\cal M}_{(0,1,\epsilon)} & =&
{\cal M}_{(1,2)}^{(0)} \qquad (N_f=0, \ {\rm no \  affine \ term})\ ,\nl
{\cal M}_{(1,0,t)}  & =&
{\cal M}_{(1,0)}^{(0)} \qquad (N_f=0, \ {\rm no \  affine \ term})\ .
\label{mcond}\earr
The unique solution is:
\beq
{\cal M}_{(1,-1,r)} = \left( \begin{array}{ccc}
0 & 1 & \epsilon \\
-1 & 2 & \epsilon \\
0 & 0 & 1 \end{array} \right) ,\
{\cal M}_{(1,1,s)} = \left( \begin{array}{ccc}
2 & 1 & -\epsilon \\
-1& 0 &  \epsilon\\
0 & 0 & 1 \end{array} \right) ,\
{\cal M}_{(1,0,t)} = \left(
\begin{array}{ccc}
1 & 0 & 0 \\
-1 & 1 & 0 \\
0 & 0 & 1 \end{array} \right) ,
\label{mtot1}\eeq
which yields the following monodromies at infinity and at the critical 
point,
\beq
{\cal M}^{(1)}_\infty = \left( \begin{array}{ccc}
-1 & 3 & \epsilon \\
0 & -1 & 0 \\
0 & 0 & 1 \end{array} \right)\ ,\qquad
{\cal M}_c^{(1)} = {\cal M}_{(1,0,t)} {\cal M}_{(1,-1,r)}=
\left( \begin{array}{ccc}
0 & 1 & \epsilon \\
-1 & 1 & 0\\
0 & 0 & 1 \end{array} \right)\ .
\label{mtot2}\eeq
Note that two further sets of identities (\ref{mcond}) can be written for the
other, ${\bf Z}_3$-symmetric quark decouplings, but they are
automatically satisfied by this solution. 
The results in this Section have been compared with the literature 
\cite{brand2}\cite{bf2}\cite{ferrari}\cite{ken}, when available;
moreover, ${\cal M}_c^{(1)}$ and ${\cal M}_{(1,1,s)}$ have been 
checked by explicitly computing the Seiberg-Witten section at 
the critical mass in Appendix B.

Next we discuss some simple physical consequences of the result 
(\ref{mtot1}),(\ref{mtot2}).
Following Ref.\cite{bf1}, we can constrain the values
of the $\{s_f\}$ charges in the BPS spectrum 
by enforcing its invariance under the spectral flow at weak coupling. 
Namely, any weak-coupling dyon state
${\bf n}=(1,n,s)$, with $n\in{\bf Z}$ and $s$ unknown, should
map into another one under the monodromy at infinity, 
corresponding to a rotation of the $\theta$-angle by three periods
\cite{ken}: this reads $(1,n,s){\cal M}^{-1}_\infty= -(1,n+3,-s-\epsilon)$. 
This flow closes on the following two sets of dyons:  
$\{(1,2n,s)\}$ and $\{(1,2n+1,-s-\epsilon)\}$,
$n\in{\bf Z}$, with $s$ still free. This is fixed to be $s=0$ 
by the further condition that the dyons with even electric charge
should survive in the $N_f=0$ theory as $m\to\infty$.
In conclusion, the complete $N_f=1$ spectrum is:
\beq
(1,2n,0)\ , \qquad (1,2n+1,-\epsilon)\ , \quad n\in{\bf Z}\ , \qquad\qquad
(N_f=1)\ .
\eeq
Thus, the pseudo-baryonic charges in the previous Eqs.(\ref{mcond}) 
are $r=s=-\epsilon$ and $t=0$.

\setcounter{footnote}{1}
Further remarks concern the behaviour at the critical point.
The Seiberg-Witten section is found to be 
$(a_D,a)(u_c)=(0,-\epsilon\ m/\sqrt{2})$
by requiring that the BPS mass of the two particles $(1,0,0)$ 
and $(1,-1,-\epsilon)$ vanishes at this point.
One verifies that this value of the section is
a non-trivial fixed point for the monodromy ${\cal M}_c$ in 
Eq.(\ref{mtot2}), as it should.\footnote{
Note, however, that ${\cal M}^3_c \neq 1$, i.e. the complete
monodromies no longer represent $SL(2,{\bf Z})$.}
Moreover, a non-vanishing value implies that 
the $W^{\pm}$ bosons ${\bf n}=(0,\pm 2,0)$ remain massive at this point,
$m_{W}=m^2$, i.e. there is no restoration of the 
non-Abelian gauge symmetry \cite{s-w}.

Finally, we determine the weak-coupling spectrum of the $N_f=2,3$ theories.
We use the corresponding monodromies at infinity computed in
Appendix A, Eqs.(\ref{m2tot},\ref{m3tot}), and we get  
the spectra of dyon charges $\{(n_m,n_e)\}$ from Ref.\cite{s-w}.
We then find the spectral flows:
\beq
\begin{array}{llcll}
{\cal M}^{(2)}_\infty : & (1,n,s_1,s_2) 
& \longrightarrow & (1,n+2,-\ep -s_1, \ep -s_2)\ , & (N_f=2);\\
{\cal M}^{(3)}_\infty : & (1,n,s_1,s_2,s_3) 
& \longrightarrow\ & (1,n+2,-\ep -s_1, \ep -s_2,\ep -s_3),& (N_f=3); \\
\ & (2,2n+1,s_1,s_2,s_3) 
& \longrightarrow\ & (2,2n+3,-2\ep -s_1, 2\ep -s_2,2\ep -s_3).& 
\end{array}
\eeq
Let us first discuss the $N_f=2$ spectrum. 
The spectral flow maps even (odd) charge
dyons within themselves; since  a tower of even dyons should survive when
$m_1$ or $m_2$ go to infinity, the set $\{(1,2n,0,0)\}$ should exist.
Then, the other set $\{(1,2n,-\epsilon,\epsilon)\}$ should also
exist by consistency of the flow. Therefore, we find that 
the even dyons are (at least) doubly degenerate with respect to 
their electric and magnetic charges.
Similar conditions of flow consistency and  ($N_f=1$) matching can be
applied to the odd dyons, which also come in pairs. The $N_f=2$ spectrum is
thus found to be:
\beq
\begin{array}{l}
(1,2n,0,0)\ , \\ (1,2n,-\ep,\ep)\ ,\end{array}
\qquad
\begin{array}{l}
(1,2n+1,-\ep,0)\ , \\ (1,2n+1,0,\ep)\ ,\end{array}
\ \qquad n\in{\bf Z}\ , \qquad\qquad (N_f=2).
\eeq
Note that the symmetry under flavour permutation is achieved up to
a gauge transformation $\ep \to -\ep$.
Actually, within our choice of patches, the decoupling of the two quarks
is associated to the monodromy decomposition
${\cal M}^{(2)}_\infty={\cal M}_{(0,1,\ep,0)}\ {\cal M}^{(0)}_\infty\ 
{\cal M}_{(0,1,0,\ep)} $.
This shows that the permutation of the quarks requires a non-trivial 
conjugation which changes the sign of $\ep$.

Next, we perform a similar analysis for the $N_f=3$ dyon spectrum 
at weak coupling. There are dyons with one unit of magnetic charge 
which turns out to be four-fold degenerate, and, moreover, 
dyons with two units of magnetic charge, which can be consistently
taken non-degenerate. One finds the spectrum:
\barr
\begin{array}{l}
(1,2n,0,0,0) \\ (1,2n,-\ep,\ep,0) \\
(1,2n,-\ep,0,\ep) \\(1,2n,0,\ep,\ep) \end{array}
\ \ \ \ &\quad&
\begin{array}{l}
(1,2n+1,-\ep,\ep,\ep) \\ (1,2n+1,0,0,\ep) \\
(1,2n,0,\ep,0) \\(1,2n,-\ep,0,0) \end{array}
\qquad \ n\in{\bf Z}\ , \qquad\quad (N_f=3)\ , \nl
&&\nl
(2,2n+1,r_1,r_2,r_3) && 
\earr
where the pseudo-baryonic charges $r_1,r_2,r_3 \neq 0$ of the 
$n_m=2$ dyons remain practically undetermined.
Let us add a few remarks:
$(i)$ The degeneracies of the $N_f=2,3$  spectra nicely match
the dimensions of their flavour multiplets  in massless limit 
\cite{s-w}\cite{bf1}\cite{brand2};
$(ii)$ The purely electric states of
quarks $(0,1,\pm\ep,\dots)$ and $W^\pm$ bosons $(0,\pm 2,0,\dots)$
are always present in these spectra: they are fixed points of the spectral
flow, as they should, and their $s_f$ 
number are not constrained, but are nevertheless known semi-classically.

In conclusion, the complete affine monodromies have been easily
found by enforcing the braiding and fusing identities.
These data of the $SU(2)$ theories allow an (almost) complete 
determination of the weak coupling spectra and their behaviour
when one or more quarks decouple.
The BPS states at strong coupling are a subsets of the weak coupling 
spectra, with unchanged quantum numbers, which can be determined
by the stability analysis of Ref.\cite{bf2}.


\section{The Analytic Isomonodromy Problem}

In the following, we prove explicitly the isomonodromy of the
Seiberg-Witten section in the simplest non-trivial case of
three singularities at finite $u$, corresponding to $N_f=1$.
The original approach to the isomonodromic problem \cite{yoshida} is
based on the study of the Fuchsian second-order differential 
equations, which can describe holomorphic sections with 
two-dimensional monodromies.
We shall apply this framework to the second-order equation 
obeyed by $( da_D/du, da/du )$:
its components are given by the period integrals of the
first and second Abelian differentials \cite{s-w}, 
$\Pi_1 = \oint dx/y $, $\ \Pi_2 = \oint x dx/y $, respectively,
where $y(x)$ is the elliptic curve (\ref{el}).
They satisfy a system of first-order differential equations, which 
are called the Picard-Fuchs equations:
\begin{eqnarray}
\frac{d\Pi_1}{du} & = & \frac{p_{11}(u)}{\Delta(u)} \Pi_1 + 
\frac{p_{12}(u)}{\Delta(u)}  \Pi_2 \ ,\nonumber \\
\frac{d\Pi_2}{du} & = & \frac{p_{21}(u)}{\Delta(u)} \Pi_1 +
\frac{p_{22}(u)}{\Delta(u)}  \Pi_2 \ ,
\label{pf}
\end{eqnarray}
where the $p_{ij} (u)$ are polynomials in the coefficients of the
elliptic curve and their derivatives \cite{yang}. 
Their explicit form in the $N_f=1$ case is, in the notation of 
the previous Section,
\begin{eqnarray}
p_{11} (u) &=& -p_{22}(u) = - 4  u^2 + 4 u m^2 + 6 m \ , \nonumber \\
p_{12} (u) & = & 6 u - 8 m^2 \ , \nonumber \\
p_{21} (u) & = & - 16 m u + 16 m^3 + 18 \ , \nonumber \\
\Delta(u) & = & 8 \left( u^3 - m^2 u^2 - 9 m u + 8 m^3 + 
\frac{27}{4}\right) \nonumber \\
& = & 8 ( u - a_1 ) ( u - a_2 ) ( u - a_3 )  \ .
\end{eqnarray}
Note that we introduced the notation $a_1,a_2,a_3$
for the three singularities in the moduli space.
It is convenient to fix two of them, 
say $a_1$ and $a_2$, to the points $ ( 0, 1 )$ by using the $SL(2,{\bf C})$
transformation $ u \rightarrow z = ( u - a_1 ) / ( a_2-a_1 )$.

The system (\ref{pf}) is equivalent to the following Fuchsian differential
equation in the variable $z$:
\beq
\left[ \frac{d^2}{dz^2} - q(z)  \right] y(z) = 0 \ , \quad
\qquad y(z) =p(z)\ \frac{da}{dz} \ ,
\eeq
where the ``potential'' $q(z)$ is,
\barr
q(z) & = & -\frac{1}{4 z^2} - \frac{1}{4 {(z-1)}^2} - \frac{1}{4
         {(z-\xi)}^2} + \frac{3}{4{(z-\eta)}^2} \nl 
& \ & + \frac{\beta_1}{z} + \frac{\beta_2}{z-1} + \frac{\beta_3}{z-\xi} 
    + \frac{\beta_4}{z-\eta} \ ,\nl
{\rm with\ }&\ &\xi \equiv { a_3 - a_1 \over a_2 - a_1}\ , \qquad 
\eta \equiv { a_4 - a_1\over a_2 - a_1}\ ,\quad a_4\equiv\frac{4}{3} m^2 \ .
\label{fuch}\earr
This equation has four regular singularities in $0, 1, \xi$ and $\infty$. 
Moreover, one finds another one at $z=\eta$, but this
does not actually correspond to a singularity of the solution $y(z)$. 
This is the so-called apparent singularity, a characteristic feature of
isomonodromy \cite{yoshida}: actually, a simple counting argument shows
that the Fuchsian equation with only physical singularities would not
have enough parameters for allowing the monodromy matrices to be 
independent of the singularity positions.
The effective coupling constant,
\beq 
\tau (u) = \frac{da_D}{da} = \frac{\theta_{\rm eff}(u)}{\pi} + i
\frac{8\pi}{g^2_{\rm eff}(u)} \ ,
\eeq
is given by the ratio of the two independent solutions of the Fuchsian
equation; moreover, its Schwarzian is proportional to the potential:
\beq 
q(z) = - \frac{1}{2} \{ \tau, z \} \ ,\qquad\quad
\{ \tau, z \} = \frac{ \tau^{'''}}{\tau^{'}} - \frac{3}{2} 
{(\frac{\tau^{''}}{\tau^{'}})}^2 \ .
\eeq
A simple consequence of these formulae is that
$ \tau' (\eta) = 0$, namely the apparent singularity is actually
a saddle point. 

The coefficients $\beta_i$ in (\ref{fuch}) are called the accessory 
parameters; their expressions for the Seiberg-Witten solution
can be obtained from the previous formulae:
\begin{eqnarray}
\beta_1(\xi,\eta) & = & - \frac{13}{36} - \frac{13}{36} \frac{1}{\xi}
+ \frac{5}{12} \frac{1}{\eta} + \frac{1}{12} \frac{\eta}{\xi}
\ ,\nonumber \\
\beta_2(\xi,\eta) & = & \frac{13}{36} - \frac{13}{36} \frac{1}{\xi-1}
+ \frac{5}{12} \frac{1}{\eta-1} - \frac{1}{12} \frac{\eta-1}{\xi-1}
\ ,\nonumber \\
\beta_3(\xi,\eta) & = & \frac{13}{36} \frac{1}{\xi} + \frac{13}{36} 
\frac{1}{\xi-1} + \frac{5}{12} \frac{1}{\eta-\xi} + \frac{1}{12} 
\frac{\eta-1}{\xi-1} - \frac{1}{12} \frac{\eta}{\xi}
\ , \nonumber \\
\beta_4(\xi,\eta) & = & - \frac{5}{12} \left( \frac{1}{\eta} +
\frac{1}{\eta-1} + \frac{1}{\eta-\xi} \right) \ .
\label{cond}
\end{eqnarray}
Moreover, the movable and apparent singularities, respectively
$\xi$ and $\eta$, are not independent in this solution, which has a 
single parameter $m$. Their relation can be obtained by analysing 
$\Delta (a_4)$ in (\ref{delta}) as a
function of the singularities $a_i$, together with the derivatives 
$\Delta ' (a_4)$ and $\Delta '' (a_4)$ . By comparing these 
polynomials, we find the relation:
\beq 
( \eta^2 - \xi )^2 - 4 \eta^2 ( \eta-1 )( \eta-\xi ) = 0 \ .
\label{pr}
\eeq

Let us now discuss the conditions for isomonodromy of 
the solution $y(z)$ or, equivalently, of $\tau(z)$.
In the general Riemann-Hilbert problem, the accessory parameters
$\beta_i(\xi,\eta)$ are not known, because the behaviour of $y(z)$ 
near the singularities only determines the second-order poles
in the Fuchsian equation. Moreover, the motion of the apparent
singularity $\eta$ as a function of $\xi$ is also unknown.
The conditions of isomonodromy provide a sufficient, 
although rather involved, set of differential equations for these unknowns,
which we shall briefly review \cite{yoshida}.
Afterwards, we shall check that these conditions are satisfied 
by the Seiberg-Witten solution.

The monodromy transformation of $\tau (z)$ does not depend on $\xi$ (or
equivalently on $m$); thus, the quantities $d\tau/d\xi$ and $d\tau/dz$
must have the same monodromy, and their ratio must be a meromorphic
function $A(z)$. This implies an auxiliary equation for $y\sim da/dz $:
\beq 
\frac{d\tau}{d\xi} = A(z) \frac{d\tau}{dz}  \ \longrightarrow \ \ 
\frac{dy}{d\xi} = A(z) \frac{dy}{dz} + B(z) y \ ,
\label{aux}\eeq
Next, we require the compatibility of this equation with the Fuchsian 
(\ref{fuch}): this determines, 
\beq
A(z) = \frac{(\eta-\xi)}{\xi(\xi-1)} \frac{z(z-1)}{(z-\eta)}\ , 
\qquad
B(z) = - \frac{1}{2} \frac{dA}{dz} \ .
\label{aux2}\eeq
Moreover, it yields four first-order differential equations 
for the $\beta_i$ in the $\eta$ and $\xi$ variables and a
relation between $\beta_4$ and $d \eta / d \xi$. 
These differential equations admit three first integrals: 
one of these amounts to the condition for a regular singularity at 
infinity,
\beq 
\sum_{i=1}^4 \beta_i = 0 \ .
\label{int1}\eeq
Another integral determines the second-order pole at infinity,
\beq 
\beta_2 + \beta_3 \xi + \beta_4 \eta = - \frac{1}{4} \ , 
\label{int2}\eeq
and the last one forbids a logarithmic behaviour of the solution near
the apparent singularity $z=\eta$. This condition implies:
\beq 
\beta^2_4 = - \frac{1}{4} \frac{1}{\eta^2} -  
\frac{1}{4(\eta-1)^2} - \frac{1}{4 (\eta-\xi)^2} + \frac{\beta_1}{\eta} +
\frac{\beta_2}{\eta - 1} + \frac{\beta_3}{\eta-\xi} \ .
\label{int3}\eeq 
These three algebraic conditions can be used to reduce the system of
four differential isomonodromy conditions to a single non-linear 
differential equation for $\eta(\xi)$, which is the Painlev\`e VI 
equation,
\begin{eqnarray}
 \frac{d^2\eta}{d\xi^2}& = & \frac{1}{2} \left( \frac{1}{\eta} + 
\frac{1}{\eta-1} + \frac{1}{\eta-\xi} \right) 
\left( \frac{d\eta}{d\xi} \right)^2
- \left( \frac{1}{\xi} + \frac{1}{\xi-1} + \frac{1}{\eta-\xi} \right) 
\frac{d\eta}{d\xi} \nonumber \\
& + &  \frac{1}{2} \frac{\eta (\eta-1)}{\xi (\xi-1)(\eta-\xi)} \ . 
\end{eqnarray} 
Note that the parameters of this equation have been set to 
the specific values of this problem.

We now verify that the $(N_f=1)$ Seiberg-Witten solution satisfies these
analytic conditions of isomonodromy.
Indeed, the coefficients $\beta_i(\xi,\eta)$ (\ref{cond}) 
deduced from the Picard-Fuchs equations can be shown to fulfil 
the first integrals (\ref{int1},\ref{int2},\ref{int3}); 
in the third equation, one should use the polynomial
relation for $\eta(\xi)$ (\ref{pr}).
Furthermore, this relation is also found to be a rather remarkable
complete integral of the Painlev\`e equation.
This concludes the proof of isomonodromy.

As discussed in the Introduction, this property is rather 
intuitive, because the Seiberg-Witten section is specified by an
elliptic curve (\ref{el}), whose parameters can be continuously deformed.
Nevertheless, it is nice to have a complete analytic proof.
Moreover, the first-order differential equation in the mass parameter
(\ref{aux},\ref{aux2}) is a new result of this Section: 
this is an auxiliary equation
for the Picard-Fuchs system (\ref{pf}) which expresses the
isomonodromy property; similar equations also exist for the Seiberg-Witten 
solution with general gauge group and may have important consequences
(see Ref.\cite{solit} in this respect).


\section{Isomonodromy and Conformal Blocks}

In Section $2$, we have shown that the Seiberg-Witten holomorphic
sections support braiding and fusing operations which are similar
to those found by Moore and Seiberg for the conformal 
blocks of rational conformal field theories \cite{mose}.
In this Section, we try to express directly the Seiberg-Witten
sections as the blocks of a suitable conformal theory.
Clearly, the conformal blocks are meant to be an operator representation
for the holomorphic sections - the two-dimensional fields do not carry
any immediate physical meaning.

Is there a general correspondence between isomonodromy and conformal
symmetry? On one hand, all the conformal blocks yield
isomonodromic holomorphic sections. This is a consequence of:
$(i)$ the locality of the operator product expansion of the conformal
fields, and $(ii)$ the (generalized) Fuchsian differential equations 
satisfied by the blocks, which originate from the chiral algebra 
in the rational conformal field theory.
On the other hand, isomonodromic sections might be non-conformal
invariant, yet being analytic: an example is given by some
correlators of the two-dimensional Ising model off-criticality 
\cite{iso}. 
However, there is more to say on this subject; in Appendix C, 
we recall some known results on the relations among isomonodromy,
integrability and conformal symmetry \cite{iso}\cite{kohno}. 
For the sake of simplicity, let us postpone this general discussion 
and proceed to present an explicit correspondence.

Let us try to express $( da_D / du, da / du )$ in the simplest cases of
two singularities at finite $u$. These are the $N_f=0$ section and the
$N_f=1,2,3$ ones at the mass values where $(N_f+1)$ singularities merge
in a maximal superconformal point $(N_f,1)$ \cite{douglas} (these 
are computed in Appendix B). 
We compare them with the four-point conformal
blocks of the Virasoro minimal models \cite{cft},
\beq 
{\cal F}_p (z) = \langle \phi_1 (z) \phi_2 (0) \phi_3 (1) \phi_4
(\infty) \rangle_p \ , \qquad  p = 1,2\ , 
\eeq
where $z = (u-a_1)/(a_2 -a_1)$ is the usual rescaled variable
$(\R (a_2-a_1)=(a_2-a_1)<0)$; the
dimensions of the fields and central charge are:
\beq 
c(p,p') = 1 - 6 \frac{{(p-p')}^2}{pp'} \ , \qquad\qquad
h_{r,s} = \frac{{( rp' - s p )}^2 - {(p-p')}^2}{4p p'}  \ .
\label{kac}\eeq
The two-dimensional monodromies imply that the field $\phi_1 (z)$ has
two-dimensional operator-product expansion with the other fields: 
this identifies $\phi_1 = \phi_{1,2}$, i.e.  
$(r,s) = (1,2)$ in the Kac table (\ref{kac});
the other fields should form a closed operator-product subalgebra
with $\phi_{1,2}$, and then belong to the set $\{ (1,n),\ n =1,2,3,..\}$.
For example, 
\beq 
\phi_{1,2}(z) \phi_{1,n} (0) = \frac{1}{z^{h_{1,2}+h_{1,n}}} \left(
z^{h_{1,n-1}}\phi_{1,n-1} + z^{h_{1,n+1}}\phi_{1,n+1} \right) 
+\quad {\rm regular}\ .
\label{ope12}\eeq
This operator-product expansion should reproduce the singular behaviours 
of the Seiberg-Witten section, which are logarithmic at infinity,
\beq 
\phi_1 (z) \phi_4 (\infty) \sim z^{-1/2} \left( 1 +
\log (z) \right) \ +\ \dots \ ,\eeq
and at $z=1$,
\beq 
\phi_1 (z) \phi_3(1) \sim 1 + \log (z-1) \ +\ \dots\ .
\label{plog}\eeq
At $(z=0)$, the behaviour is again logarithmic for the $N_f=0$ section, 
while it is a power law for $N_f=1,2,3$, due to the superconformal 
symmetry, with exponents $( \pm 1/6, \pm 1/4, \pm 1/3 )$, respectively 
(see Appendix B for more details).

The logarithmic behaviour is rather unusual in conformal field theory,
but it is nevertheless possible for the non-unitary ``logarithmic''
theories $c(1,p')$. As explained in Ref. \cite{gurarie}, 
a logarithmic operator-product expansion is
obtained when the two fields appearing in the r.h.s. of (\ref{ope12}) 
have the same dimension:
\beq 
h_{1,n-1} = h_{1,n+1} \qquad \longrightarrow \qquad c = c(1,n). 
\eeq
Actually, the theory $c(1,p')$ may contain more than one pair of
degenerate dimensions, but only the pair $(\phi_{1,p'-1} ,
\phi_{1,p'+1})$ is obtained by the $\phi_{1,2}$ operator product expansion.
Furthermore, a purely logarithmic behaviour (\ref{plog}) would require
$h_{1,2} + h_{1,p'} = h_{1,p'-1}$, which is impossible. 
Nonetheless, this can be cured by adding a power law prefactor 
to the conformal blocks. 
Next, we compare the explicit form of the Seiberg-Witten section with
the general expression of the minimal conformal blocks given by the
Dotsenko-Fateev Coulomb gas. Actually, in Section $3$ of 
Ref.\cite{dotsenko}, an Euler-type integral representation of 
the four-point block $\langle
\phi_{n,m} \phi_{1,2} \phi_{1,2} \phi_{n,m}\rangle$ is given, together with
its Hypergeometric expressions.

Let us first consider the $N_f=0$ theory. From Ref.\cite{s-w}, we obtain,
\begin{eqnarray}
\frac{da_D}{dz}& = & -i\ F \left( \frac{1}{2}, \frac{1}{2}, 1 ; z \right)\ , 
\nonumber \\
\frac{da}{dz}& = & -\frac{1}{2} (1-z)^{-\frac{1}{2}} \ 
F \left( \frac{1}{2}, \frac{1}{2}, 1 ; \frac{1}{1-z} \right)\ , 
\end{eqnarray}
where $z=(1-u)/2$.
By matching $da_D/dz$ with the parameters in the Dotsenko-Fateev
expression, we find the unique identification with the following blocks 
of the $c(1,2)=-2$ minimal model:
\beq 
\left( \frac{da_D}{dz}, \frac{da}{dz} \right) 
 \sim \lim_{z_4 \to\infty}
{ \langle \phi_{12} (z) \phi_{12} (0) \phi_{12} (1) \phi_{12} (z_4)
\rangle_p \over \left( z (1-z) z_4 \right)^{1/4} } \ ,\ \ c =
c(1,2)=-2\ \ \ (N_f=0).\eeq
Note that the prefactors cancel out in the representation for the effective
coupling constant $\tau=da_D/ da$ as a ratio of two blocks.

The Seiberg-Witten sections of the $N_f=1,2,3$ theories at their
critical mass values, given in Appendix B, 
Eqs.(\ref{n1basis},\ref{n1sol},\ref{n23sol}),
can be similarly represented by
the blocks involving a pair of $\phi_{1,2}$ fields and another
$\phi_{1,n}$ pair. One finds:
\begin{eqnarray}
N_f = 1:\ \left( \frac{da_D}{dz}, \frac{da}{dz} \right) 
& \sim &
{ \langle \phi_{13} (z) \phi_{13} (0) \phi_{12} (1) \phi_{12} (\infty)
\rangle_p \over z^{1/2} (1-z)^{1/3} } \ ,\ \ \ c = c(1,3) = -7 \ ,
\nonumber \\ 
N_f = 2:\ \left( \frac{da_D}{dz}, \frac{da}{dz} \right) 
& \sim &
{ \langle \phi_{14} (z) \phi_{14} (0) \phi_{12} (1) \phi_{12} (\infty)
\rangle_p \over z^{7/8} (1-z)^{3/8} } \ , \ \ \ c =
c(1,4) = - \frac{25}{2} \ ,\nonumber \\ 
N_f = 3:\ \left( \frac{da_D}{dz}, \frac{da}{dz} \right) 
& \sim &
{ \langle \phi_{16} (z) \phi_{16} (0) \phi_{12} (1) \phi_{12} (\infty)
\rangle_p \over z^{7/4} (1-z)^{5/12} } \ ,\ \ \ c = c(1,6) = -24 .\ 
\end{eqnarray}
Therefore, we were able to relate the analytic properties of
the Seiberg-Witten theory with those of the conformal field
theory. The Moore-Seiberg braiding and fusing operators acting on the
conformal fields can be applied in this case for describing the
isomonodromic properties, thus improving the analysis of Section $2$. 

On the other hand, this equivalence of isomonodromy and conformal 
symmetry is only valid
for the case of three singularities, i.e. the four-point blocks.
We have not been able to represent the general $N_f=1$ section as 
a five-point blocks of the logarithmic minimal models: actually, 
the fusion rules of $\phi_{12}$ fields are consistent with the merging of 
singularities in critical points, and identify the fields 
entering the five-point block. However, it is impossible to match all 
the asymptotic behaviours with the corresponding operator 
product expansions.

This failure can be understood as follows.
In general, we know that the Seiberg-Witten holomorphic sections
satisfy a Fuchsian equation with $(3+N_f)$ regular singularities: the
three-singularity solution is necessarily an Hypergeometric function,
which transforms covariantly under the $SL(2,{\bf C})$ regular conformal
transformations of the $u$-plane, and can be represented by correlator of
primary conformal fields. The general $(3+N_f)$-singularity solutions
may not have such covariance: actually, their explicit expressions 
or integral representations \cite{lag}\cite{bf2},
and their differential equations show that these sections are not
$SL(2,{\bf C})$ covariant. For example, we can consider 
the representation using
the Hypergeometric of argument the modular invariant function
$j(\tau ( u))$ \cite{brand}, and find that the mapping $j(u)$ violates 
$SL(2,{\bf C})$ covariance.
Furthermore, the section of the $SU(3)$ gauge theory has been
expressed in terms of the generalised Hypergeometric function of two variables 
$F_4$ in Ref.\cite{lerche}; this is also not conformal covariant 
because its integral representation 
is not of the Euler type as the conformal blocks \cite{dotsenko}. 

A possible breaking of conformal symmetry while keeping analyticity could
be obtained by correlators of non-primary fields, which depend analytically
on dimensionful parameters. This program can be practical only if we
can characterize these new fields by other transformation properties which
replace the $SL(2,{\bf C})$ covariance.
In this respect, our approach can be related to other attempts in the
literature \cite{moro} to relate the Seiberg-Witten solutions to 
two-dimensional topological field theories \cite{topft}. 
Actually these theories, as well as the matrix models and
(quantum) Liouville theory \cite{df}, are further known isomonodromic systems
which do not necessarily have conformal symmetry.

Finally, let us discuss the general relation between isomonodromic
sections and conformal blocks given by the theory of holonomic quantum 
fields of the Refs.\cite{iso}. These Authors 
have analysed the $k$-dimensional isomonodromic sections in the context
of the Schlesinger system of equations, which generalizes the $k=2$
analysis of the Fuchsian equation in Section $3$ (see Appendix C for
more details). They have
expressed the general isomonodromic solution as a correlator of
$k$-component Weyl fermions $\Psi^{\alpha} (z)$,
$\overline{\Psi}_\be (z)$, $\al,\be=1,\dots,k$, and the ``twist fields''
$V(a_i)$, $i = 1, .., n$, which belong to a $c=k$ conformal field
theory. The monodromy around the singularity $z=a_i$ is described
by the operator-product expansion $\Psi^\alpha (z) V(a_i)$,
and the prescribed monodromy matrix 
$(M_i)^\beta_\alpha$ determines the following form of the twist field:
\beq 
V(a_i) = : {\cal P} \exp\left( - \int_{\infty}^{a_i} \ dy (\log
M_i)^\beta_\alpha \ \overline{\Psi}_\beta (y) \Psi^\alpha (y)
\right) : \ .
\label{twistf}\eeq
For general, non diagonal $M_i$ matrices, this generalized vertex
operator is only formally defined as a path-ordered $({\cal P})$ series;
therefore, its conformal covariance and locality are not ensured.
We conclude that this approach is also affected by the 
difficulty of defining the field operators encountered before.
Again, this general representation by
twist fields correlators can be useful once we 
understand the specific covariance of these fields. 

\vspace{1cm}

\noindent{\large\bf Acknowledgements}

We would like to thank Kenichi Konishi for many useful discussions.
This work is supported in part by the European Community Program
``Training and Mobility of Researchers'' FMRX-CT96-0012.

\vspace{.5cm}
\noindent{\large \bf Note Added}

Section $2$ may have some overlap with the recent papers in the Refs.
\cite{bf2}\cite{ken}; we would like to note that 
our results were obtained independently,
but the writing was delayed by other obligations.

\appendix


\section{Monodromies of the $N_f=2,3$ Theories}

\begin{figure}
\epsfxsize=14cm 
\centerline{\epsfbox{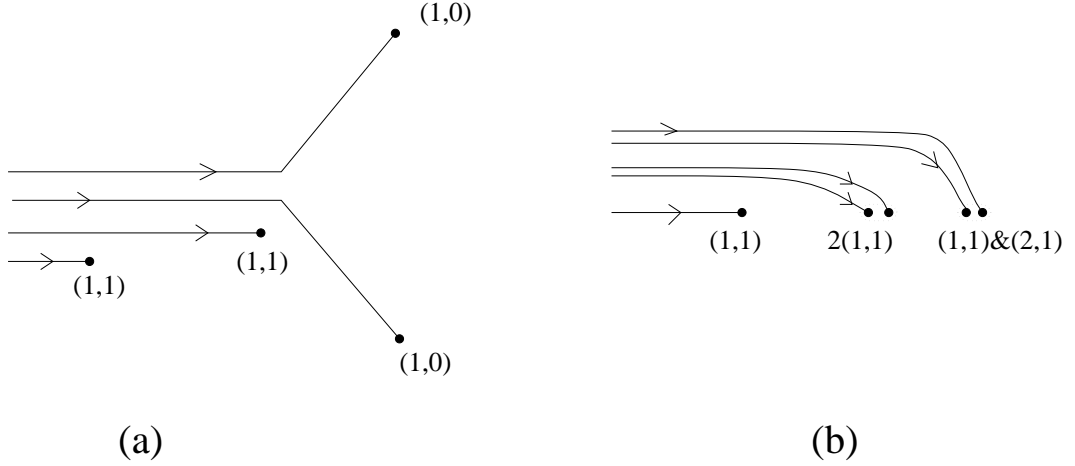}}
\caption{Singularities and cuts for the theories: (a) $N_f = 2,  m_1 =
  m_2 > 0$; (b), $N_f = 3$, $ m_1 = 0$, $m_2 = m_3 = m > 0$.}
\label{fig8}
\end{figure}

We now generalise the results of Section $2.3$ to the cases $N_f > 1$. 
As for $N_f = 1$, we vary the values of the the masses and follow the
consequent motion of the singularities in the moduli space. 
We then use the braiding and fusing identities as equations 
for determining the complete monodromy matrices.

\underline{$N_f = 2$ case} \\
When $m_1 = m_2 = 0$, there is a double singularity at 
$ u =  \Lambda_{2}$ which represents two massless monopoles $(1,0)$,
and a double singularity at $u = - \Lambda_{2}$ of two
massless dyons $(1,1)$. Their monodromies are represented, respectively, by 
${\cal M}_{(1,0)},{\cal M}_{(1,0)}'$,${\cal M}_{(1,1)}$,
${\cal M}_{(1,1)}'$, such that 
${\cal M}_{\infty}^{(2)} = {\cal M}_{(1,1)} {\cal M}_{(1,1)}'
{\cal M}_{(1,0)} {\cal M}_{(1,0)}' $.
If we switch on $m_2$ and make it growing,
the two monopole singularities separate each other; one goes in  the upper
half plane and the other in the lower half plane (see Fig.(9 a)). 
The two dyon singularities separate along the real axis, one of them
passes between the monopole pair and then goes to infinity: this 
becomes the massless quark singularity by the effect of the braiding
with the lower monopole. Next, the quark cut should be rotated of 
$ -\pi$, such that it points to Re(z) = $+ \infty$ and disappears after
decoupling, leaving the $N_f = 1$ singularities in strong coupling region: 
this cut rotation requires a further conjugation of the
upper (1,0) monopole with the quark, yielding $(1,0) \longrightarrow (1,-1)$.

The complete $N_f = 2$ monodromies are $(4 \times 4)$ matrices,
with unknown affine terms, which are plugged in the previous braid
relations,
\barr
{\cal M}_{(1,0)}^{-1}\ {\cal M}_{(1,1)}'\  
{\cal M}_{(1,0)}  = {\cal M}_{(0,1,0,\epsilon)} \equiv {\cal M}_{quark\ 2}
\ ,\nl
{\cal M}_{quark\ 2}\  {\cal M}_{(1,0)}'\  
{\cal M}_{quark\ 2}^{-1} = 
{\cal M}_{(1,-1,-\epsilon)}^{(N_f = 1)} \ ,
\earr
and are matched with the $N_f = 1$ monodromies in Eq.(\ref{mtot1}),
\beq
{\cal M}_{(1,1)}^{(N_f = 2)} = {\cal M}_{(1,1,-\epsilon)}^{(N_f = 1)}\ .
\eeq
We find the unique solution: 
\barr
{\cal M}_{(1,1)} = 
\left(
\begin{array}{cc|cc}
2 & 1 & -\epsilon & 0   \\
-1 & 0 & \epsilon & 0   \\ \hline
\multicolumn{2}{c|}{\bf 0} & \multicolumn{2}{c}{\bf 1}  \\
\end{array}
\right)\ ,\quad
{\cal M}_{(1,1)}' = \left(
\begin{array}{cc|cc}
2 & 1 & 0 &   \epsilon   \\
-1 & 0 & 0 & -\epsilon   \\ \hline
\multicolumn{2}{c|}{\bf 0} & \multicolumn{2}{c}{\bf 1}  \\
\end{array}
\right) \ ,
\nl
{\cal M}_{(1,0)} = 
\left(
\begin{array}{cc|cc}
1 & 0 & 0 & 0   \\
-1 & 1 & 0 & 0   \\ \hline
\multicolumn{2}{c|}{\bf 0} & \multicolumn{2}{c}{\bf 1}  \\
\end{array}
\right)\ ,\quad
{\cal M}_{(1,0)}' = 
\left(
\begin{array}{cc|cc}
1 & 0 & 0 & 0   \\
-1 & 1 & \epsilon & -\epsilon   \\ \hline
\multicolumn{2}{c|}{\bf 0} & \multicolumn{2}{c}{\bf 1}  \\
\end{array}
\right) \ ,
\label{a3}\earr
where $\epsilon$ is the same sign ambiguity of the $N_f = 1$ case. 
From these it is simple to find the expressions of the monodromies 
at infinity and at the (2,1) critical point, occurring for
$m_1=m_2=\sqrt{2}$:
\barr
{\cal M}_{\infty}^{(2)} &= &
\left(
\begin{array}{cc|cc}
-1 & 2 & \epsilon & -\epsilon   \\
0 & -1 & 0 & 0   \\ \hline
\multicolumn{2}{c|}{\bf 0} & \multicolumn{2}{c}{\bf 1}  \\
\end{array}
\right) \ ,\nl
{\cal M}_{c}^{(2)} = {\cal M}_{(1,1)}' {\cal M}_{(1,0)} {\cal M}_{(1,0)}'
&= &\left(
\begin{array}{cc|cc}
 0 & 1 & \epsilon & 0   \\
-1 & 0 & 0 & -\epsilon  \\ \hline
\multicolumn{2}{c|}{\bf 0} & \multicolumn{2}{c}{\bf 1}  
\end{array}
\right)\ .
\label{m2tot}\earr

\underline{$N_f = 3$ case} \\
When $m_1 = m_2 = m_3 = 0$, there are two singularities on the real
axis of  the moduli
space, corresponding to four massless dyons $(1,1)$ and a
(2,1) massless dyon, respectively. 
Their monodromies are denoted by ${\cal M}_{(1,1)},{\cal M}_{(1,1)}'$,
${\cal M}_{(1,1)}^{\prime \prime}$,
${\cal M}_{(1,1)}^{\prime \prime \prime}$, such that
${\cal M}_{\infty}^{(3)} = {\cal M}_{(1,1)} {\cal M}_{(1,1)}' 
{\cal M}_{(1,1)}^{\prime \prime} 
{\cal M}_{(1,1)}^{\prime \prime \prime}
{\cal M}_{(2,1)}$.
When $m_3$ acquires a real positive value, a double singularity 2 (1,1) and
the (2,1) one approach each other and make a superconformal
point. Afterwards, they split again with different quantum numbers, i.e.
the one on the right has become the quark singularity. By allowing an
imaginary part to $m_3$, one can resolve this motion and actually find a
double braiding of (2,1) around the pair 2(1,1), such that 
$(2,1) \longrightarrow (1,0)$. Therefore, the braid relation and
matching conditions to impose are: 
\barr
{\cal M}_{(1,1)}^{\prime \prime}\ {\cal M}_{(1,1)}^{\prime \prime \prime}
\ {\cal M}_{(2,1)}
\left({\cal M}_{(1,1)}^{\prime \prime}
{\cal M}_{(1,1)}^{\prime \prime \prime} \right)^{-1} = 
{\cal M}_{(0,1,0,0,\epsilon)} = {\cal M}_{quark\ 3} \ ,
\nl
{\cal M}_{quark\ 3}\ {\cal M}_{(1,1)}^{\prime \prime}\
{\cal M}_{(1,1)}^{\prime \prime \prime}\ 
{\cal M}_{quark\ 3}^{-1} = 
{\cal M}_{(1,0)}^{(N_f = 2)} {\cal M}_{(1,0)}^{\prime\ (N_f = 2)}\ ,
\nl
{\cal M}_{(1,1)} {\cal M}_{(1,1)}' = 
{\cal M}_{(1,1)}^{(N_f = 2)}{\cal M}_{(1,1)}^{\prime\ (N_f = 2)}\ .
\label{a5}\earr
Other braid relations can be obtained from another pattern of
decoupling: by letting $m_2 = m_3 = m$ real and positive, a (1,1)
singularity, followed by a 2(1,1) pair, move toward the (2,1) one (Fig
9 b); the (1,1) and (2,1) fuse into a
critical point, and then separate, going in the
upper and lower half-plane, respectively. It is simple to see that, after the
separation, the singularity going downwards is a monopole (1,0), while
the other one is the dyon (1,1).
The 2(1,1) singularity proceeds to the right and passes between these two;  
the braiding rules show that
this becomes a double quark singularity. By decoupling these quarks
to the right, there remain the $N_f = 1$ singularities and we find:
\beq
{\cal M}_{\infty}^{(3)} = {\cal M}_{\infty}^{(1)} {\cal M}_{quark\ 2}\ 
{\cal M}_{quark\ 3}\ . 
\eeq
The corresponding braid relations and matching conditions are,
\barr
{\cal M}_{(1,1)}^{\prime \prime \prime}{\cal M}_{(2,1)} 
\left( {\cal M}_{(1,1)}^{\prime \prime \prime} \right)^{-1} =
{\cal M}_{(1,0)}^{(N_f = 1)}\ ,
\nl
{\cal M}_{quark\ 2}\ {\cal M}_{quark\ 3}\ 
{\cal M}_{(1,1)}^{\prime \prime \prime} 
\left({\cal M}_{quark\ 2}\ {\cal M}_{quark\ 3}\right)^{-1} = 
{\cal M}_{(1,-1)}^{(N_f = 1)}\ ,
\nl
{\cal M}_{(1,1)} = {\cal M}_{(1,1)}^{(N_f = 1)}\ .
\label{a7}\earr
The equation (\ref{a5}) and (\ref{a7}) are sufficient to determine all the
affine terms for the monodromy matrices; the result is:
\barr
{\cal M}_{(1,1)} = 
\left(
\begin{array}{cc|ccc}
2 & 1 & -\epsilon & 0 & 0  \\
-1 & 0 & \epsilon & 0 & 0   \\ \hline
\multicolumn{2}{c|}{\bf 0} & \multicolumn{3}{c}{\bf 1}  \\
\end{array}
\right) &,&
{\cal M}_{(1,1)}^{\prime} = 
\left(
\begin{array}{cc|ccc}
 2 & 1 & 0 &  \epsilon & 0  \\
-1 & 0 & 0 & -\epsilon & 0   \\ \hline
\multicolumn{2}{c|}{\bf 0} & \multicolumn{3}{c}{\bf 1}  \\
\end{array}
\right)\ , 
\nl
{\cal M}_{(1,1)}^{\prime \prime} = 
\left(
\begin{array}{cc|ccc}
 2 & 1  & 0 & 0 & \epsilon \\
-1 & 0 & 0 & 0  & -\epsilon  \\ \hline
\multicolumn{2}{c|}{\bf 0} & \multicolumn{3}{c}{\bf 1}  \\
\end{array}
\right) &,&
{\cal M}_{(1,1)}^{\prime \prime \prime} = 
\left(
\begin{array}{cc|ccc}
2 & 1 & -\epsilon & \epsilon & \epsilon  \\
-1 & 0 & \epsilon & -\epsilon & -\epsilon   \\ \hline
\multicolumn{2}{c|}{\bf 0} & \multicolumn{3}{c}{\bf 1}  \\
\end{array}
\right)\ ,
\nl
{\cal M}_{(2,1)} &= &
\left(
\begin{array}{cc|ccc}
3 & 1 & -\epsilon & \epsilon & \epsilon  \\
-4 & -1 & 2 \epsilon & - 2 \epsilon & - 2 \epsilon   \\ \hline
\multicolumn{2}{c|}{\bf 0} & \multicolumn{3}{c}{\bf 1}  \\
\end{array}
\right) .
\label{nf3m}
\earr
From these, we calculate the monodromies at the infinity and at
the critical point (3,1):
\barr
{\cal M}_{\infty}^{(3)} = 
\left(
\begin{array}{cc|ccc}
-1 & 1 & \epsilon & -\epsilon & -\epsilon  \\
0 & -1 & 0 & 0 & 0   \\ \hline
\multicolumn{2}{c|}{\bf 0} & \multicolumn{3}{c}{\bf 1}  \\
\end{array}
\right)\ ,
\nl
{\cal M}_{c}^{(3)} = 
\left(
\begin{array}{cc|ccc}
0 & 1 & \epsilon & 0 & 0  \\
-1 & -1 & 0 & -\epsilon &  -\epsilon   \\ \hline
\multicolumn{2}{c|}{\bf 0} & \multicolumn{3}{c}{\bf 1}  \\
\end{array}
\right)\ ,
\label{m3tot}\earr
which are used in the text to derive the $s_i$ quantum numbers.


\section{Explicit Solutions at the Critical Masses}

In the recent literature, there have appeared rather explicit formulae
for the Seiberg-Witten sections of the general massive
$SU(2)$ theories with $N_f=1,2,3$: one expression uses the Elliptic
functions, after putting the elliptic curves in the Weierstrass 
normal form \cite{lag}\cite{bf2}; another one is given by the 
Hypergemetric function with argument the modular invariant 
function $j(\tau(u))$ \cite{brand}.
These expressions depend on the moduli space coordinate $u$ 
in a rather indirect way, such that the analytic 
continuations are rather difficult in general.
Hereafter, we shall describe the simple cases in which the Seiberg-Witten
section can be directly expressed in the $u$ variable 
as an integral of the Hypergeometric function, such that the
derivation of the complete monodromies is rather straightforward.

These cases are characterized by the presence of only three singularities
in the moduli space, when $(N_f+1)$ singularities fuse into the $(N_f,1)$
superconformal points (in the notation of Ref.\cite{douglas}).
By extending the argument of Section $3$, we can argue that the
derivative $(da_D/du,da/du)$ satisfies a Fuchsian differential 
equation in the $u$ variable with three singularities, 
whose solution is necessarily Hypergeometric. 
We use the $SL(2,{\bf C})$-reduced variable $z=(u-a_1)/(a_2-a_1)$,
with $\R (a_2-a_1)=(a_2-a_1) <0$, 
and put the singularities in $z=0,1,\infty$.
The logarithmic behaviours occur at $z=1,\infty$, and the
power-law singularity is found at the critical point $z=0$.
Its characteristic indices can be deduced by the corresponding 
monodromies matrices, which represent the $SL(2,{\bf Z})$ group, 
in particular by the diagonalizable elements: $S$ ($S^2 = -1$) and $ST$ 
($(ST)^3=1$) or combinations of them.
Form the analysis of Section $2$ and Appendix A, we know that the critical
monodromies are $(-TS)$, $S$ and $(ST)$ for $N_f=1,2$ and $3$, respectively.
The corresponding indices $(a,b,c)$ of the Hypergeometric function 
$F(a,b,c;z)$ are
($1/6,1/6,1/3$),($1/4,1/4,1/2$) and ($1/3,1/3,2/3$), respectively.

To start with, let us describe the case $N_f=1$. 
We use the following bases of solutions:
\begin{eqnarray}
y_1(\infty) & = & k_1 (-z)^{(-\frac{1}{2})} F\left( \frac{1}{3}, 
\frac{2}{3}, 1; \frac{1}{z} \right) \ ,\nonumber \\
y_2(\infty) & = & k_1 (-z)^{(-\frac{1}{2})} \left[  log(-z)  
F\left( \frac{1}{3}, \frac{2}{3}, 1 ;\frac{1}{z} \right) \right. \nl
&\  & +\left. 
\sum_{n=0}^{+\infty} \frac{(\frac{1}{3})_n (\frac{2}{3})_n}{(n!)^2} 
\left( 2 \psi(n+1) - \psi\left(n+\frac{1}{3}\right) - 
\psi\left(n+\frac{2}{3}\right) \right) 
\left(\frac{1}{z}\right)^n \right] \ ,\nonumber \\
y_1(1) & = & k_1 (-z)^{(-\frac{1}{6})} 
F\left( \frac{1}{3}, \frac{1}{3}, 1 ; 1 - z\right) \ ,\nonumber \\
y_2(1) & = & k_1 (-z)^{(-\frac{1}{6})} \left[  log(1-z)  
F\left( \frac{1}{3}, \frac{1}{3}, 1 ; 1 - z \right) \right.\nonumber \\
& \ & +\left. 
2 \sum_{n=0}^{+\infty} \frac{(\frac{1}{3})_n (\frac{1}{3})_n}{(n!)^2} 
\left(\psi\left(n+\frac{1}{3}\right) - \psi(n+1) \right) 
\left( 1 - z \right)^n \right] \ , \nonumber \\
y_1(0) & = & k_1 (-z)^{(-\frac{1}{6})}
F\left( \frac{1}{3}, \frac{1}{3}, \frac{2}{3} ; z \right)
\ , \nonumber \\
y_2(0) & = & k_1 (-z)^{(-\frac{1}{6})}
F\left( \frac{2}{3}, \frac{2}{3}, \frac{4}{3} ; z \right)\ ,
\label{n1basis}\end{eqnarray}
which have simple monodromy transformations around each point.
For example, around $z=\infty$ we have:
\beq 
y_1(\infty) \ \rightarrow \ - y_1(\infty) \qquad\qquad
y_2(\infty) \ \rightarrow \ - y_2(\infty) - 2i \pi\ y_1(\infty) \ .
\eeq
The three solutions in (\ref{n1basis}) are related by the 
linear transformations:
\begin{eqnarray}
y_1 (\infty) & = & \frac{1}{2 ( \psi(\frac{1}{3}) - \psi(\frac{2}{3}) )}
\left( \frac{\Gamma(\frac{2}{3})^2}{\Gamma(\frac{4}{3})}
y_2(0) - \frac{\Gamma(\frac{1}{3})^2}{\Gamma(\frac{2}{3})} y_1(0)
 \right) \ , \nonumber \\ 
y_2(\infty) & = &  \frac{1}{2} \left(  
\frac{\Gamma(\frac{2}{3})^2}{\Gamma(\frac{4}{3})}
y_2(0) + \frac{\Gamma(\frac{1}{3})^2}{\Gamma(\frac{2}{3})} y_1(0)
 \right) \ , \nl
y_1(\infty) & = & e^{\mp i \frac{\pi}{3}} y_1 (1) -
\frac{\sqrt{3}}{2\pi} \left( 1 - e^{\mp i \frac{\pi}{3}}\right) 
y_2 (1)\ , \nonumber \\
y_2(\infty) & = & - \frac{\pi}{\sqrt{3}} e^{\mp i \frac{\pi}{3}} y_1 (1)
- \frac{1}{2} \left( 1 + e^{\mp i \frac{\pi}{3}} \right) y_2 (1)\ . 
\label{basetr}\end{eqnarray}
In the last transformation, the upper (lower) sign refers to  
the case $\I z>0$ ($ \I z <0$).

The expressions for $da/dz$ and $da_D/dz$ are linear
combinations of this basis which satisfy the asymptotics and
reproduce the expected monodromies: the asymptotics 
$da/du=\ep/2\sqrt{2u} + O(u^{-3/2})$ at 
$z\propto -u\to -\infty$ identifies
$da/dz=y_1(\infty)$ with $k_1=\ep \sqrt{3}m_c/2\sqrt{2}$, 
upon using the critical value of the mass $m_c=3/2$ and $(a_1-a_2)=27/4$ 
obtained from $\Delta(u)$in Eq.(\ref{delta}); moreover, the sign 
$\epsilon=\pm 1$ accounts for the choice of the square-root
for $u\sim 2a^2$. The linear combination for $da_D/dz$
is determined by the monodromies at infinity (\ref{minf1}) and at
$z=0$: in the latter case, one applies the transformation of 
basis (\ref{basetr}). The result is:
\barr
{da_D\over dz} &= & \frac{1}{2}\ y_1 (\infty) + \frac{3i}{2\pi}\ 
y_2 (\infty) \ ,
\qquad\quad \left(k_1=\ep {\sqrt{3}\ m_c\over 2\sqrt{2}}\right) \nl       
{da\over dz} &= & y_1(\infty) \ , 
\label{n1sol}\earr
Next, $a$ and $a_D$ can be represented as contour 
integrals in the $u$-plane of the Hypergeometric functions; 
the integration constant for $a$ is determined by its asymptotic
behaviour at infinity. The other boundary condition is chosen 
at the critical point: $a_D(0)=0$ for any $N_f$.
This is consistent with the choice
of patches in Section $2$ and Appendix A, because $a_D(0)=0$ is
left invariant by the action of all the critical monodromies.
Therefore, we define,
\beq
a = \int_{-\infty}^z \ dz \ \frac{da}{dz} \ ,\qquad
a_D =  \int^z_0 \ dz \ \frac{da_D}{dz}\ .
\label{adef}\eeq 
The complete monodromy matrices for ($a, a_D$) 
can be now computed as follows: one performs the analytic continuation
of the integrand and deforms the integration contour accordingly.
For the monodromy at infinity, we find,
\begin{eqnarray} 
z\sim\infty :\qquad a_D & \rightarrow & 
-a_D + 3 a - 2 \int^0_{-\infty} \frac{da_D}{dz} \ , \nl
a & \rightarrow & - a \ . 
\end{eqnarray}
The affine terms are tabulated integrals of the Hypergeometric \cite{grad}:
\begin{eqnarray} \!\!\!\!\!\!\!
\int^0_{-\infty} dz  \frac{da}{dz} & = & k_1 
\frac{\sqrt{3}}{2\pi} \left[  
\frac{\Gamma(\frac{1}{3})^2}{\Gamma(\frac{2}{3})} 
\int^0_{-\infty} dz {(-z)}^{(-\frac{1}{6})} F\left( \frac{1}{3}, 
\frac{1}{3}, \frac{2}{3} ; z \right) - \right. \nonumber \\
& \ & -\left. \frac{\Gamma(\frac{2}{3})^2}{\Gamma(\frac{4}{3})} 
\int^0_{-\infty} dz {(-z)}^{(-\frac{1}{6})} F\left( \frac{2}{3},
\frac{2}{3}, \frac{4}{3} ; z \right) \right] = -  \frac{ 2 k_1}{\sqrt{3}}
=-\ep {m_c\over \sqrt{2}}\ ;
\end{eqnarray}
and similarly,
\beq
\int^0_{-\infty} dz \frac{da_D}{dz} =-{\ep\over 2} {m_c\over \sqrt{2}}\ .
\eeq
Therefore, the monodromy at infinity is found to agree with
the result of the algebraic approach (\ref{mtot2}) of Section $2.3$. 
This is also the case for the monodromy around the critical point, 
which is similarly found by using (\ref{basetr}). 
The monodromy around the third point $z=1$ is slightly more
subtle: it depends on the base point, since the cut from $z=0$ passes
by this point. Upon using the corresponding two transformations
of basis in (\ref{basetr}), we find the monodromies,
\begin{eqnarray}
z\sim 1\ ,&\ & \I z>0 :\nl
a_D & \rightarrow & -a_D +4a - 4\int^0_{-\infty} dz \frac{da}{dz} + 
2\int^1_0 dz \frac{d(a_D- 2a)}{dz} = -a_D +4a \ ,
\nonumber \\
a & \rightarrow & - a_D +3a - 2\int^0_{-\infty} dz \frac{da}{dz} + 
\int^1_0 dz \frac{d(a_D - 2a)}{dz} =-a_D +3a \ ;\nl
z\sim 1\ ,&\ &\I z<0 :\nl
a_D & \rightarrow & 2a_D + a - \int^0_{-\infty} dz \frac{da}{dz}
- \int_0^1 dz\frac{d(a+ a_D)}{dz} = 2a_D + a -\ep {m_c\over\sqrt{2}}\ ,\nl
a & \rightarrow & - a_D +  \int^0_{-\infty} dz \frac{da}{dz} + 
\int^1_0 dz  \frac{d(a +a_D)}{dz} = -a_D + \ep {m_c\over\sqrt{2}} \ .
\end{eqnarray}
The affine part can be evaluated thanks to:
\beq
\int^1_0 dz \frac{d(a_D - 2a)}{dz} = -2\ep {m_c\over\sqrt{2}}\ ,
\qquad\qquad
\int^1_0 dz \frac{d(a_D +a)}{dz} = 2\ep {m_c\over\sqrt{2}} \ .
\eeq
In conclusion, the second of these transformations checks again 
the result of Section $2.3$, Eq.(\ref{mtot1}).

By using the same methods for the $N_f=2$ and $N_f=3$ theories, we have
found the explicit solutions for the mass values  
$m'_c=m_1=m_2=\sqrt{2}$, implying the critical point $(2,1)$, 
and the values $m^{\prime\prime}_c=m_1=m_2=m_3=1$ for the $(3,1)$ point, 
respectively \cite{douglas}. These read:
\begin{eqnarray}
N_f &=& 2: \nl
\frac{da}{dz} & = & k_2 (-z)^{(-\frac{1}{2})} 
F \left( \frac{1}{4}, \frac{3}{4}, 1 ;\frac{1}{z} \right) \ , \nl
\frac{da_D}{dz} & = & k_2 \frac{i}{\pi} (-z)^{(-\frac{1}{2})} \left[
log(-z) F\left( \frac{1}{4}, \frac{3}{4}, 1; \frac{1}{z} \right) \right.\nl
& \ & + \left. 
\sum_{n=0}^{+\infty} \frac{(\frac{1}{4})_n (\frac{3}{4})_n}{(n!)^2} 
\left(2 \psi(n+1) - \psi\left(n+\frac{1}{4}\right) - 
\psi\left(n+\frac{3}{4}\right) \right) \left(
\frac{1}{z} \right)^n  \right] \ ,\nl
N_f & =& 3: \nl
\frac{da}{dz} & = & k_3 (-z)^{(-\frac{1}{2})} 
F \left( \frac{1}{6}, \frac{5}{6}, 1 ;\frac{1}{z} \right) \ ,\nonumber \\
\frac{da_D}{dz} & = & -\ k_3 \frac{1}{2} (-z)^{(-\frac{1}{2})} 
F \left( \frac{1}{6},\frac{5}{6}, 1 ; \frac{1}{z} \right) 
+ k_3 \frac{i}{2 \pi} (-z)^{(-\frac{1}{2})} 
\left[ log(-z) F\left( \frac{1}{6}, \frac{5}{6}, 1; \frac{1}{z} \right) 
\right. \nonumber \\
& \ & +\left. 
\sum_{n=0}^{+\infty} \frac{(\frac{1}{6})_n (\frac{5}{6})_n}{(n!)^2} 
\left( 2 \psi(n+1) - \psi\left(n+\frac{1}{6}\right) - 
\psi\left(n+\frac{5}{6}\right) \right) \left(\frac{1}{z} \right)^n\right]\ .
\label{n23sol}\end{eqnarray}
The proportionality constants are found to be:
\beq
k_2= \ep {m'_c\over \sqrt{2}}\ ,\qquad\qquad
k_3= \ep {3\sqrt{3}\over 4}\ {m^{\prime\prime}_c\over \sqrt{2}}\ .
\eeq
The section $(a_D,a)$ is again defined by the contour integrals 
in Eq.(\ref{adef}) and the monodromies are found by the same approach.
The results check the algebraic methods of Appendix A,
Eqs.(\ref{a3}) and (\ref{nf3m}), in the cases of equal masses.

\section{Isomonodromy, Integrability and Con\-for\-mal 
Symmetry}

In this Appendix, we recall some useful results in the literature
\cite{iso}\cite{kohno} 
which may clarify the general relations among these subjects.
An isomonodromic holomorphic section $Y^\al(z;a_i)$ with 
$n$ singular points $(a_1,\dots,a_n)$ and $k$-dimensional 
monodromies, $\al=1,\dots,k$, can be associated to a flat
connection.
Consider the configuration space, which is the tensor space 
${\bf C}^{n+1}$ for $z$ and the singularity positions $\{a_i\}$,
and represent the holomorphic section as a path-ordered 
exponential of a connection ${\cal A}$ in configuration space,
\beq
Y(z;a_i) = {\cal P}\ \exp\left(\ \oint_\ga\ {\cal A}\ \right) \ ,
\label{pord}\eeq
where the open path $\ga$ runs from infinity to the point
$(z;a_1,\dots,a_n)$.
This identification makes sense only if the path-ordered expression
in the r.h.s. is independent of shape of the path, and thus depends on
its end-point only. This implies that ${\cal A}$ is a flat connection, 
i.e. satisfies the zero-curvature condition,
\beq
d\ {\cal A} +{\cal A} \wedge {\cal A} = 0 \ ,
\label{cur}\eeq
where,
\beq
d={\partial\over\partial z} dz + \sum_{i=1}^n
{\partial\over\partial a_i} da_i \ ,\qquad {\rm and} \qquad
{\cal A}= \sum_{i=1}^n\ A_i(a_j) \ d \log(z-a_i) \ , 
\label{coord}\eeq
are the component forms of the differential and the connection,
respectively.
The identification (\ref{pord}) implies that $Y$ satisfies the system of
first-order differential equations:
\beq
\left( d \ -\ {\cal A} \right) Y =0 \ .
\label{difeq}\eeq
The zero-curvature condition yields the integrability conditions
for this system, which should be satisfied by the $(k\times k)$
matrix functions $A_i ( a_j)$.

A monodromy transformation on $Y^\al(z;a_i)$ is obtained by joining a 
closed path $\ga'_k$ to $\ga$ in (\ref{pord}), which extends in the
$z$ sub-plane of ${\bf C}^{n+1}$ and encircles the singularity $z=a_k$.
The flatness of the connection implies the invariance
under deformations of $\ga'_k$ off the $z$-plane in ${\bf C}^{n+1}$,
which amount to displacements of the other singularities:
this is the isomonodromy property.

This general formulation of the isomonodromic problem 
makes manifest a number of interesting relations:

{\bf I.} The equations (\ref{difeq}) and (\ref{cur}) 
can be analysed in components and compared with those
of Section $3$ based on the Fuchsian equation. 
The $dz$ component of (\ref{difeq}) is the $k$-dimensional 
differential system,
\beq 
\frac{d}{dz} \ Y^\alpha \ = \ \sum_{i=1}^n 
\frac{\left(A_i\right)^\al_\be}{z-a_i} \ Y^\beta \ ,
\eeq
which can be considered as a defining property for $Y$, analogue to the
Picard-Fuchs equations of Section $3$. The $da_i$ components give
auxiliary parametric equations,
\beq 
\frac{d}{da_i} \ Y \ = \ - \frac{A_i}{z-a_i} \ Y \ ,
\eeq
which enforce isomonodromy. These are the analogues of the 
$\xi$ (mass) equation (\ref{aux}) found in the Fuchsian theory.
Moreover, the zero curvature condition gives a set of non-linear compatibility
conditions for the $A_i(a_j)$ known as the Schlesinger equations,
\begin{eqnarray}
\frac{\partial}{\partial a_i} \ A_i &= & \sum_{j=1,\ j\neq i}^n \ 
\frac{ \left[ A_i,A_j \right] }{a_i -a_j} \ ,\nonumber \\
\frac{\partial}{\partial a_i} \ A_j &= & -  
\frac{ \left[ A_i, A_j \right]}{a_i -a_j} \ ,\qquad \ i \neq j \ ; 
\end{eqnarray}
these generalize the $\beta_i$ conditions and the Painlev\`e equation 
of Section $3$.

As mentioned in Section $4$, a formal solution  of the
system (\ref{difeq}) has been given 
in terms of the analytic correlators of the $k$-component Weyl
fermions $\Psi^{\alpha} (z), \overline{\Psi}_{\alpha} (z)$, 
$\alpha = 1, ..., k $, and the twist fields $V(a_i)$, 
which belong to a $c=k$ conformal field theory \cite{iso}. 
It has the following form:
\beq
Y^\alpha_\beta ( z, z_0 ; a_i ) = (z_0 - z) \frac{ \langle
\overline{\Psi}_\beta (z_0) \Psi^\alpha (z) V_1 ( a_1) .... V_N (a_n)
\rangle }{ \langle V_1 ( a_1) .... V_N (a_n) \rangle }\ , 
\eeq
where $z_0$ is a base point and the additional index $\beta$
specifies the choice of boundary conditions for the solutions; 
the form of the twist fields is given in Eq.(\ref{twistf}).
The derivation of this representation follows the same steps of the
analysis in Section $4$: one compares the operator-product expansion
$\Psi(z)\ V(a_i)$ around each singularity $a_i$ with the given monodromy 
data, and then determines the field $V(a_i)$ by formal
integration of this local expansion. 

\bigskip

{\bf II.} 
A general formulation of integrable systems in two dimensions
can be given in terms of a flat connection on an certain space;
therefore, we can consider the zero-curvature condition (\ref{cur}) arising
in the isomonodromy problem as the defining equations of the associated
integrable system. This is a heuristic argument for
showing that isomonodromy and integrability have a common origin.

\bigskip

{\bf III.} 
The correlators of Rational Conformal Field Theories often satisfy the
Knizh\-nik-Zamolodchikov equations, which follows from the 
symmetry under an affine Lie algebra $\widehat{\cal G}$ 
(implying conformal symmetry) \cite{cft}.
These equations can also be written in terms 
of a flat connection in configuration space, the Kohno connection 
\cite{kohno}, and correspond to a special case of the previous 
isomonodromy problem.
This discussion also applies to the general Rational
Conformal Field Theories without affine algebra, 
including the Virasoro minimal models of Section $4$, because they
can be obtained by the former theories {\it via} the coset construction, 
which replace ${\cal G}$ with ${\cal G}/{\cal H}$ \cite{cft}.

As seen in Section $4$, the $n$-singularity section $Y(z,a_i)$ should
be compared with a $(n+1)$-point correlator 
$\langle g_{n+1} (z) g_1 (a_1) g_n(a_n) \rangle$, where the point $z$ 
is considered on the same footing as the $a_i$ ones, e.g. $z=a_{n+1}$.
The conformal fields $g_i(a_i)$ carry a unitary representation $R_i$
of the Lie group ${\cal G}$ of dimension $d_i$, such that the monodromy 
problem is defined in the tensor space $\otimes_{i=1}^{n+1} R_i$ of dimension 
$k = \sum_{i=1}^{n+1} \ d_i$. The Lie group generators acting on $R_i$
are represented by $t^a_i$, satisfy $[t_i^a, t^b_i] = f^{abc} t^c_i$,
and $ [t^a_i, t^b_j]=0$ for $i\neq j$. 
Kohno has introduced the connection $\Omega$, which is a symmetrized
version of $\cal A$:
\begin{eqnarray}
 {\cal A} &\equiv & {\cal A}_{n+1}  =  \sum_{i=1}^{n} \ A_{n+1,i} \
d \log (a_{n+1} - a_i ) \ \longrightarrow \nonumber \\
\Omega & = & \frac{1}{2} \sum_{j=1}^{n+1} \ {\cal A}_j = \frac{1}{2}
\sum_{j=1}^{n+1} \sum_{i=1,i\neq j}^{n+1} \ A_{ji}\ d \log (a_j - a_i)\ .
\end{eqnarray} 
The component matrices are 
$A_{ij} = (1/\kappa)\ \sum^{d_A}_{a=1} \ t^a_i \otimes t_j^a$,
with $\kappa$ a parameter and $d_A$ the dimension of the adjoint
representation. These matrices are independent on the
singularity positions $a_i$, namely they are a special case of the Schlesinger
ones in (\ref{coord}). The usual equation $(d-\Omega) Y = 0$ read, in
components,
\beq 
\kappa \frac{\partial}{\partial a_i} \ Y (a_1,\dots,a_{n+1})
\ = \ \sum_{j=1, j\neq i}^{n+1}\ 
\frac{t^a_i\otimes t^a_j}{a_i-a_j} \ Y (a_1,\dots,a_{n+1})\ ,
\eeq
which are indeed the Knizhnik-Zamolodchikov equations. 
Moreover, the corresponding zero-curvature condition (\ref{cur}) 
reduces to the following algebraic conditions, known as the
``infinitesimal braid relations'' \cite{kohno},
\beq
\left[ A_{ij}, A_{ik} + A_{jk} \right] = 0 \ ,\qquad
\left[A_{ij} , A_{kl} \right] = 0 \ ,\qquad\quad i \neq j \neq k \neq l\ , 
\eeq
which are automatically satisfied due to the underlying group structure.
In conclusion, the Rational Conformal Field Theories represent special 
isomonodromy problems with underlying affine and conformal symmetries.

\def\NP{{\it Nucl. Phys.\ }}
\def\PRL{{\it Phys. Rev. Lett.\ }}
\def\PL{{\it Phys. Lett.\ }}
\def\PR{{\it Phys. Rev.\ }}
\def\CMP{{\it Comm. Math. Phys.\ }}
\def\IJMP{{\it Int. J. Mod. Phys.\ }}
\def\MPL{{\it Mod. Phys. Lett.\ }}
\def\RMP{{\it Rev. Mod. Phys.\ }}
\def\AP{{\it Ann. Phys. (NY)\ }}

\end{document}